\newcommand*{\ATLASLATEXPATH}{} 

\pdfoutput=1

\documentclass[cernpreprint,texlive=2014,txfonts,USenglish]{atlasdoc}

\usepackage[backend=bibtex]{\ATLASLATEXPATH atlaspackage}

\usepackage{\ATLASLATEXPATH atlasbiblatex}

\usepackage{\ATLASLATEXPATH atlascontribute}

\usepackage{\ATLASLATEXPATH atlasphysics}
\usepackage{\ATLASLATEXPATH atlasbsm}
\usepackage{\ATLASLATEXPATH atlasmisc}

\usepackage{tikz}
\usetikzlibrary{shapes.callouts,decorations.pathmorphing,snakes}
\def\Put(#1,#2)#3{\leavevmode\makebox(0,0){\put(#1,#2){#3}}}

\graphicspath{{logos/}{figures/}}

\usepackage{DVMET_paper-defs}

\usepackage{multirow}

\usepackage{caption}

\AtlasTitle{Search for long-lived, massive particles in events with displaced vertices and missing transverse momentum in $\sqrt{s} = 13$~\TeV~$pp$ collisions with the ATLAS detector}

\author{The ATLAS Collaboration}

\PreprintIdNumber{CERN-EP-2017-202}

\AtlasJournal{Phys.\ Rev.\ D.}

\AtlasAbstract{%
A search for long-lived, massive particles predicted by many theories beyond the Standard Model is presented. The search targets final states with large missing transverse momentum and at least one high-mass displaced vertex with five or more tracks, and uses \lumi\ of $\sqrt{s} = 13$~\TeV~$pp$ collision data collected by the ATLAS detector at the LHC. The observed yield is consistent with the expected background. The results are used to extract 95\% CL exclusion limits on the production of long-lived gluinos with masses up to 2.37~\TeV~and lifetimes of $\mathcal{O}(10^{-2})\text{--}\mathcal{O}(10)$~ns in a simplified model inspired by Split Supersymmetry.
}

\hypersetup{pdftitle={ATLAS document},pdfauthor={The ATLAS Collaboration}}

\begin{document}

\maketitle

\tableofcontents

\clearpage

\section{Introduction}
\label{sec:intro}

The lack of explanation for the dark matter observed in the universe~\cite{Bertone:2004pz}, the gauge hierarchy problem~\mbox{\cite{Weinberg:1975gm,Susskind:1978ms}}, and the lack of exact gauge coupling unification at high energies~\cite{Dimopoulos:1981yj} all indicate that the Standard Model (SM) is incomplete and needs to be extended. Many attractive extensions of the SM have been proposed, but decades of searches have set severe constraints on the masses of promptly decaying particles predicted by these models. Searches targeting the more challenging experimental signatures of new long-lived particles (LLPs) have therefore become increasingly important and must be pursued at the Large Hadron Collider (LHC).

A number of beyond-SM (BSM) models predict the existence of massive particles with lifetimes in the picoseconds to nanoseconds range. Many of these particles would decay in the inner tracker volume of the experiments at the LHC. The decay products of such particles often contain several electrically charged particles, which can be reconstructed as tracks. If the LLP decays within the tracking volume but at a discernible distance from the interaction point (IP) of the incoming beams, a displaced vertex can be reconstructed by using dedicated tracking and vertexing techniques.

\begin{figure}[!bp]
\begin{center}
\includegraphics[width=0.3\textwidth]{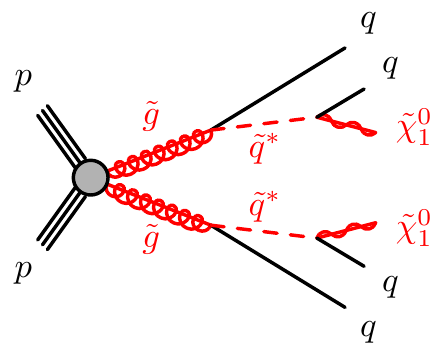}
\caption{\label{fig:feynDiag} Diagram showing pair-production of gluinos decaying through $\gluino \rightarrow \qqbar\neutralino$ via a virtual squark $\squark^*$. In Split SUSY scenarios, because of the very large squark mass, the gluinos are long-lived enough to hadronize into $R$-hadrons that can give rise to displaced vertices when they decay.
}
\end{center}
\end{figure}

There are various mechanisms by which particles obtain significant lifetimes in BSM theories. The decays of such particles can be suppressed in so-called Hidden Valley models \cite{Strassler:2006im} where large barrier potentials reduce the rate of kinematically allowed decays. Long-lived particles also appear in models with small couplings, such as those often found in $R$-parity-violating supersymmetry (SUSY)~\cite{Barbier:2004ez,Allanach:2006st}. Finally, decays via a highly virtual intermediate state also result in long lifetimes, as is the case for a simplified model inspired by Split SUSY~\cite{ArkaniHamed:2004fb,Giudice:2004tc} used as a benchmark model for the search presented here. In this model, the supersymmetric partner of the gluon, the gluino ($\gluino$), is kinematically accessible at LHC energies while the SUSY partner particles of the quarks, the squarks (\squark), have masses that are several orders of magnitude larger. \Fig{\ref{fig:feynDiag}} shows pair-production of gluinos decaying to two quarks and the lightest supersymmetric particle (LSP), assumed to be the lightest neutralino (\neutralino). The $\gluino \rightarrow \qqbar\neutralino$ decay is suppressed as it proceeds via a highly virtual squark. Depending on the scale of the squark mass, the gluino lifetime can be picoseconds or longer, which is above the hadronization time scale. Therefore, the long-lived gluino, which transforms as a color octet, is expected to hadronize with SM particles and form a bound color-singlet state known as an $R$-hadron~\cite{Fairbairn:2006gg}.

This search utilizes the ATLAS detector and attempts to reconstruct the decays of massive $R$-hadrons as displaced vertices (DVs). The analysis searches for LLP decays occurring $\mathcal{O}(1\text{--}100)$~mm from the reconstructed primary vertex (PV), and is sensitive to decays of both electrically charged and neutral states emerging from the PV. The analysis targets final states with at least one DV with a high reconstructed mass and a large track multiplicity in events with large missing transverse momentum \MET. This analysis builds on that of \Ref{\cite{Aad:2015rba}} where the ATLAS Collaboration set limits on such processes using 8~\TeV\ $pp$ collisions from the LHC. In Run 2 of the LHC starting in 2015, the increased center-of-mass energy of $\sqrt{s} = 13$~\TeV\ gives significant increases in the production cross sections of heavy particles, providing extended mass sensitivity compared to previous searches. Decays of new, long-lived particles have been searched for in a variety of experimental settings. These include studies by ATLAS~\cite{Aad:2012kw, Aad:2013oua, Aad:2014gfa, Aad:2013yna, Aad:2015asa, Aad:2015uaa, Aad:2011zb, Aad:2012zx,SUSY-2016-02,SUSY-2016-03}, CMS~\cite{Chatrchyan:2012jna, Khachatryan:2014mea, CMS:2014hka, CMS:2014wda, Khachatryan:2015wka, Khachatryan:2016unx,CMS-EXO-15-010,CMS-EXO-11-035}, LHCb~\cite{Aaij:2014nma, Aaij:2015tna, Aaij:2016xmb, Aaij:2016qsm}, CDF~\cite{Abe:1998ee}, D0~\cite{Abazov:2006as, Abazov:2009ik}, BaBar~\cite{Lees:2015rxq}, Belle~\cite{Liventsev:2013zz} and ALEPH~\cite{Heister:2002vh}. The searches involve a range of experimental signatures, including final states with leptons, jets and combinations thereof. Dedicated techniques make use of non-pointing or delayed photons, as well as tracking, energy and timing measurements of the long-lived particle itself until it decays.

The experimental apparatus is described in~\Sect{\ref{sec:detector}}, and~\Sect{\ref{sec:datasets}} discusses the data set and simulations used for this analysis. The special reconstruction algorithms and event selection criteria are presented in \Sect{\ref{sec:recoandevsel}}. \Sect{\ref{sec:backgrounds}} discusses the sources of backgrounds relevant to this search and the methods employed to estimate the expected yields. The sensitivity to experimental and theoretical uncertainties of the analysis is described in \Sect{\ref{sec:uncertainties}}. \Sect{\ref{sec:results}} presents the results and their interpretations.

\section{ATLAS detector}
\label{sec:detector}

The ATLAS experiment~\cite{PERF-2007-01,ATLAS-TDR-19} at the LHC is a multi-purpose particle detector with a forward-backward-symmetric cylindrical geometry and a near $4\pi$ coverage in solid angle.\footnote{ATLAS uses a right-handed coordinate system with its origin at the nominal IP in the center of the detector and the $z$-axis along the beam pipe. The $x$-axis points from the IP to the center of the LHC ring, and the $y$-axis points upward. Cylindrical coordinates $(R,\phi)$ are used in the transverse plane, $\phi$ being the azimuthal angle around the beam pipe. The pseudorapidity is defined in terms of the polar angle $\theta$ as $\eta=-\ln\tan(\theta/2)$.} The detector consists of several layers of subdetectors. From the IP outwards there is an inner tracking detector (ID), electromagnetic and hadronic calorimeters, and a muon spectrometer (MS).

The ID extends from a cylindrical radius of about 33~mm to 1100~mm and to $|z|$ of about 3100~mm, and is immersed in a 2~T axial magnetic field. It provides tracking for charged particles within the pseudorapidity region $|\eta| < 2.5$. At small radii, silicon pixel layers and stereo pairs of silicon microstrip detectors provide high-resolution position measurements. The pixel system consists of four barrel layers, and three forward disks on either side of the IP. The barrel pixel layers, which are positioned at radii of 33.3~mm, 50.5~mm, 88.5~mm, and 122.5~mm are of 
particular relevance to this work. The silicon microstrip tracker (SCT) comprises four double layers in the barrel and nine forward disks on either side.  The radial position of the innermost (outermost) SCT barrel layer is 299~mm (514~mm). The final component of the ID, the transition-radiation tracker (TRT), is positioned at larger radii, with coverage up to $|\eta|=2.0$.

The calorimeter provides coverage over the range $|\eta| < 4.9$. It consists of an electromagnetic calorimeter based on lead and liquid argon with coverage for $|\eta|<3.2$ and a hadronic calorimeter. Hadronic calorimetry in the region $|\eta|<1.7$ uses steel absorbers and scintillator tiles as the active medium. Liquid-argon calorimetry with copper absorbers is used in the hadronic end-cap calorimeters, which cover the region $1.5 < |\eta| < 3.2$. A forward calorimeter using copper and tungsten absorbers with liquid argon completes the calorimeter coverage up to $|\eta| = 4.9$.

The MS consists of three large superconducting toroid systems each containing eight coils and a system of trigger and precision tracking chambers, which provide trigger and tracking capabilities in the range $|\eta| < 2.4$ and $|\eta| < 2.7$, respectively.

A two-level trigger system is used to select events~\cite{TRIG-2016-01}. The first-level trigger is implemented in custom electronics and uses information from the MS trigger chambers and the calorimeters. This is followed by a software-based high-level trigger system, which runs reconstruction algorithms similar to those used in offline reconstruction. Combined, the two levels reduce the $40$~MHz bunch-crossing rate to approximately $1$~kHz of events saved for further analysis.

\section{Data set and simulated events}
\label{sec:datasets}

The experimental data used in this paper are from proton--proton ($pp$) collisions at $\sqrt{s} = 13$~\TeV\ collected in 2016 at the LHC. After applying requirements on detector status and data quality, the integrated luminosity of the sample corresponds to \lumi. The uncertainty in the 2016 integrated luminosity is 2.2\%. It is derived, following a methodology similar to that detailed in Ref.~\cite{DAPR-2013-01}, from a calibration of the luminosity scale using $x$--$y$ beam-separation scans performed in May 2016.

This search makes use of a number of signal Monte Carlo (MC) samples to determine the efficiency for selecting signal events and the associated uncertainty. In each sample, gluinos were pair-produced in $pp$ collisions and then hadronized, forming metastable $R$-hadrons. The gluino contained in each $R$-hadron later decays to SM quarks and a neutralino as shown in \Fig{\ref{fig:feynDiag}}. The mass of the gluino ($m_{\gluino}$) in the simulated samples is between $400$ and $2000$~\GeV, its lifetime $\tau$ varies from $0.01$ to $50$~ns, 
and the neutralino mass $m_{\tilde{\chi}_{1}^{0}}$ ranges from $100~\GeV$ to $m_{\gluino}-30$~\GeV. 
To evaluate signal efficiencies for lifetimes not simulated, events in the produced samples are reweighted to different lifetimes.  
The samples were simulated with \PYTHIAV{6.428}~\cite{Sjostrand:2006za}. The AUET2B~\cite{ATL-PHYS-PUB-2011-014} set of tuned parameters for the underlying event and the CTEQ6L1~\cite{Pumplin:2002vw} parton distribution function (PDF) set are used. Dedicated routines~\cite{Farrar:1978xj,Kraan:2004tz,Fairbairn:2006gg} for hadronization of heavy colored particles were used to simulate the production of $R$-hadrons. The hadronization process primarily yields meson-like states ($\gluino \qqbar$), but baryon-like states ($\gluino qqq$) and glueball-like states ($\gluino g$) are predicted as well. Following the hadronization, approximately half of the $\gluino$-based $R$-hadrons have electric charge $Q \neq 0$, and the charges of the two $R$-hadrons produced in the event are uncorrelated. The electric charge of the $R$-hadron is determined by its SM parton content, and while $Q = -1$, $0$ and $1$ dominate, a few percent have double charge.  It is worth noting that the vertexing algorithms used in this search (see in \Sect{\ref{sec:trackingandvertexing}}) are agnostic to the electric charge of the LLP as only the decay products are reconstructed.

The cross sections are calculated at next-to-leading order (NLO) assuming a squark mass large enough to completely decouple squark contributions. The most significant contributions to the NLO QCD corrections come from soft-gluon emission of the colored particles in the initial and final states~\cite{Beenakker:1996ch,Borschensky:2014cia,Kulesza:2009kq}. The resummation of soft-gluon emission is taken into account at next-to-leading-logarithm accuracy (NLO$+$NLL)~\cite{Beenakker:1996ch,Kulesza:2008jb,Kulesza:2009kq}. The uncertainty in the cross-section predictions is defined as an envelope of the predictions resulting from different choices of PDF sets (CTEQ6.6~\cite{Nadolsky:2008zw} and MSTW2008~\cite{Martin:2009iq}) and the factorization and renormalization scales, as described in \Ref{\cite{Borschensky:2014cia}}. The nominal cross section is obtained using the midpoint of the envelope.

The ATLAS detector simulation~\cite{Aad:2010ah}
is based on \geant~\cite{Agostinelli:2002hh}, and dedicated routines are employed to simulate interactions of $R$-hadrons with matter~\cite{Kraan:2004tz,Mackeprang:2006gx,Mackeprang:2009ad}. The model used assumes an $R$-hadron--nucleon cross section of $12$~mb per nucleon for each light valence quark of the $R$-hadron. For glueball-like states ($\gluino g$), the interaction cross section is assumed to be the same as for the meson-like states ($\gluino \qqbar$). The per-parton interaction probability is roughly inversely proportional to the squared parton mass, rendering the interactions of the gluinos themselves negligible. For the glueball-like states, $g\rightarrow \qqbar$ transitions create an effective mass for the gluon similar to that of the meson-like states~\cite{Kraan:2004tz}. 

The decay of the $R$-hadron is simulated by a modified version of \PYTHIAV{6.428} and includes the three-body decay of the gluino, fragmentation of the remnants of the light-quark system, and hadronization of the decay products. In all signals considered, the kinematics of the decay products are determined primarily by the mass of the gluino and the kinematics of the $R$-hadron it is contained in.

$R$-hadron production was simulated using \PYTHIAV{6.428}; however, it is not expected to accurately model the initial-state radiation (ISR) or final-state radiation (FSR). To obtain a more accurate description of these effects, additional samples of $\gluino\gluino$ production were generated using \MGMCatNLO~2.2.3~\cite{Alwall:2014hca} and interfaced to the \PYTHIAV{8.186} parton shower model, with the A14~\cite{ATL-PHYS-PUB-2014-021} set of tuned parameters together with the NNPDF2.3LO~\cite{Ball:2012cx} PDF set. The distribution of the transverse momentum \pt\ of the $\gluino\gluino$ system simulated with \PYTHIAV{6} is reweighted to match the distribution obtained for corresponding \MGMCatNLO samples.

All MC samples include simulation of additional $pp$ interactions in the detector from the same or nearby bunch crossings, referred to as pileup. These additional inelastic $pp$ interactions that occur in the detector were generated using \PYTHIAV{8.186}~\cite{Sjostrand:2007gs} tuned with the A2 parameter set~\cite{ATL-PHYS-PUB-2012-003} and overlaid with the hard-scattering event. Simulated events are reconstructed using the same algorithms used for the collision data. 

\section{Reconstruction and event selection}
\label{sec:recoandevsel}

While the reconstruction of DV candidates makes use of the ID, the entirety of the ATLAS detector is used to reconstruct the jets and \met\ in each event, thereby providing additional discrimination between signal and background. 
Hadronic jets are reconstructed from calibrated three-dimensional topo-clusters~\cite{PERF-2014-07} using the \antikt jet clustering algorithm \cite{Cacciari:2008gp,Cacciari:2011ma} with a radius parameter of $0.4$. Jet candidates are initially calibrated assuming their energy depositions originate from electromagnetic showers, and then corrected by scaling their four-momenta to the energies of their constituent particles~\cite{PERF-2011-05,PERF-2012-01,Aaboud:2017jcu,ATL-PHYS-PUB-2015-027}. Electrons, photons, and muons are also reconstructed and calibrated, although no explicit requirements are placed on them in this search. The \met\ is calculated using all calibrated objects as well as those reconstructed tracks not associated with these objects. The latter contribution accounts for potential diffuse, low-\pt\ imbalances~\cite{PERF-2011-07,PERF-2014-04}.

\subsection{Reconstruction of displaced tracks and vertices}
\label{sec:trackingandvertexing}

In the standard ATLAS tracking algorithm~\cite{ATL-PHYS-PUB-2015-018}, triplets of hits in the pixel and/or the SCT detectors are used to seed the track finding. By adding further hits along the seed trajectories, track candidates are fitted and subsequently extrapolated into the TRT. This algorithm places constraints on the transverse and longitudinal impact parameters of track candidates with respect to the PV\footnote{The PV is required to have at least two associated tracks and satisfy $|z|<200$~mm. If several exist, the vertex with the largest $\sum (\pT^{\mathrm{track}})^2$ is selected.} ($|d_0| < 10~\mm$ and $|z_0| < 250~\mm$, respectively). These constraints result in low efficiency for reconstructing tracks originating from a DV, as such tracks typically have a larger transverse impact parameter than those emerging from the interaction point.

In order to recover tracks from DVs, an additional  \emph{large-radius tracking} (LRT) algorithm pass~\cite{ATL-PHYS-PUB-2017-014} is performed, using only hits not already associated with tracks reconstructed by the standard tracking algorithm. Requirements on the impact parameters are relaxed, allowing tracks to have $|d_0| < 300~\mm$ and $|z_0| < 1500~\mm$. Furthermore, requirements on the number of hits shared by several tracks are slightly relaxed. The tracks from the standard processing and the LRT processing are treated as a single collection in the subsequent reconstruction steps.

Tracks satisfying \mbox{$\pt>1~\gev$} are selected for the DV reconstruction. In order to remove fake tracks, a track is discarded if it simultaneously has no TRT hits and fewer than two pixel hits. Tracks with fewer than two pixel hits are therefore required to fall within the TRT acceptance of $|\eta|<2$. Tracks are also required to have \mbox{$|d_0| > 2~\mm$} in order to reject tracks that originate from the PV and from most short-lived particles, such as $b$-hadrons. This last requirement also ensures that the track from an electrically charged LLP will not be associated with the DV.

The DV reconstruction algorithm starts by finding two-track seed vertices from pairs of selected tracks. Seed vertices with a high quality of fit are retained. Both tracks of a seed vertex are required to not have hits in pixel layers at smaller radii than the seed vertex, and to have a hit in the nearest pixel or SCT layer at larger radius. If the seed vertex position is inside or within several millimetres of a tracker layer, hits of that particular layer are neither forbidden nor required.
Kinematic requirements on the direction of the vector sum of the momenta of the tracks associated with the seed vertex are applied to make sure it is consistent with the decay of a particle originating from the PV. 

At this stage, a track can be associated with multiple two-track seed vertices. In order to resolve such ambiguities, an iterative process based on the incompatibility graph approach~\cite{Das:1973} is applied. After this procedure, each track is associated with at most one seed vertex.

Multi-track DVs are then formed iteratively using the collection of seed vertices. For a given seed vertex ${\V_1}$, the algorithm finds the seed vertex ${\V_2}$ that has the smallest value of $d/\sigma_d$, where $d$ is the three-dimensional distance between ${\V_1}$ and ${\V_2}$, and $\sigma_d$ is the estimated uncertainty in $d$. If $d/\sigma_d < 3$, a single DV is formed from all the tracks of both seed vertices and the merged vertex is refitted. The merging is repeated until no other compatible seed vertices are found. Simultaneously, the significance of each track's association with its vertex is evaluated upon merging, and poorly associated tracks not satisfying additional criteria are removed before the vertex is refitted. This procedure is repeated until no other tracks fail to meet these criteria. Finally, DVs separated by less than $1~\mm$ are combined and refitted. DV candidates are only considered in this search if they fall in the fiducial volume $R = \sqrt{x^2+y^2}<300$~mm and $|z|<300$~mm.

\begin{figure}[!htbp]
\begin{center} 
\subfloat[]{	
	\label{fig:vertexingeffvsrlrt}
	\includegraphics[width=0.45\textwidth]{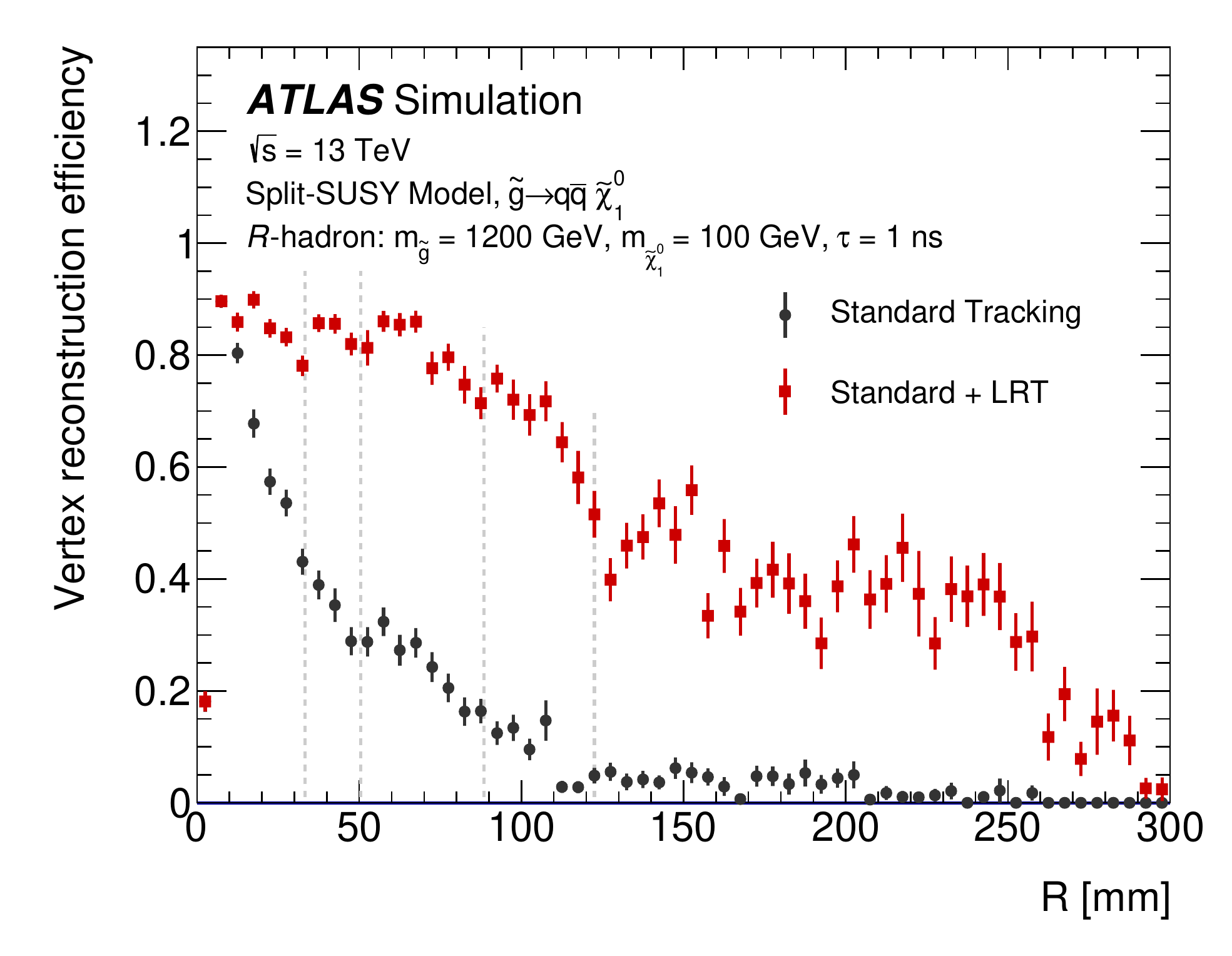}
}
\subfloat[]{
	\label{fig:vertexingeffvsr}
	\includegraphics[width=0.45\textwidth]{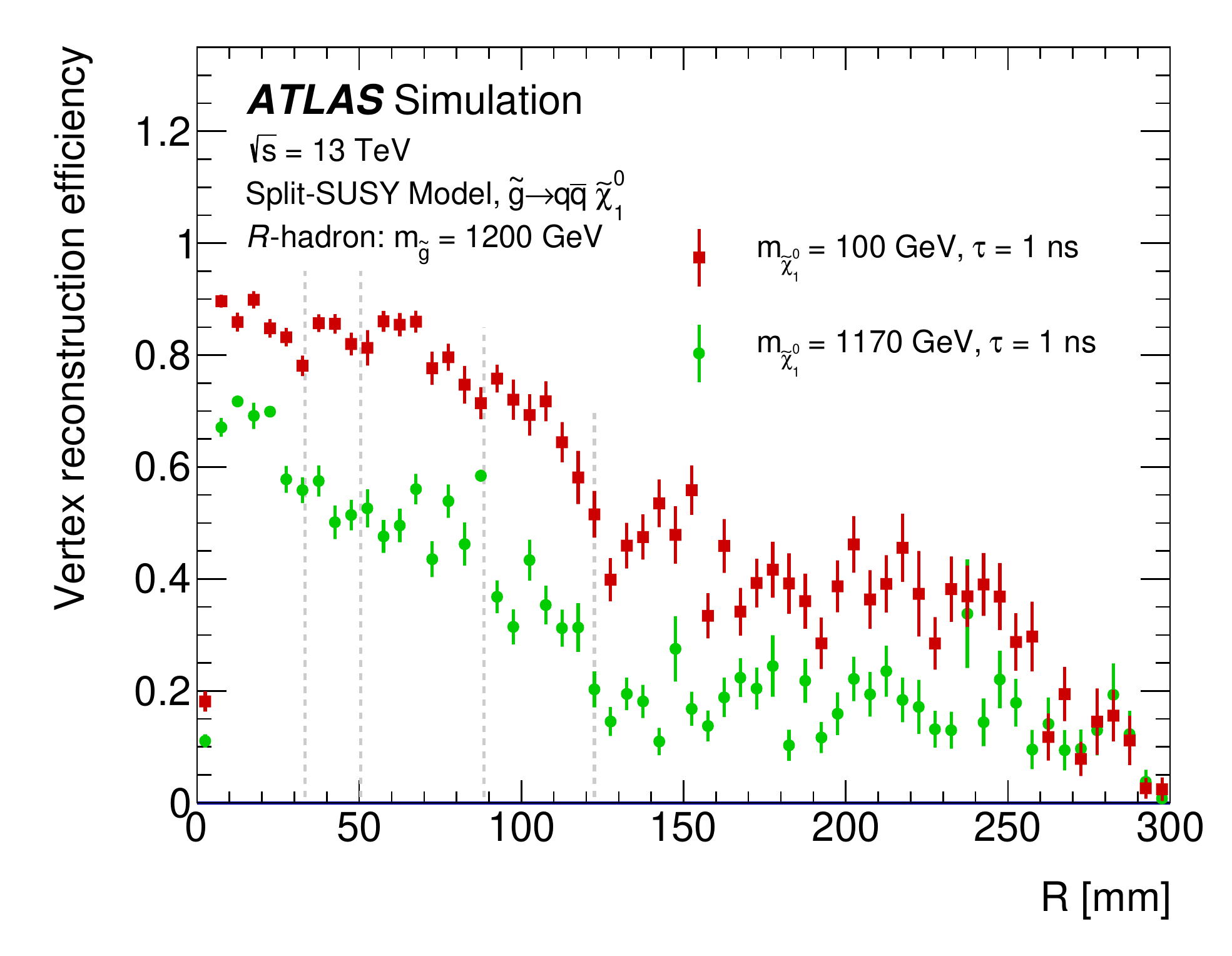}
}
\caption{\label{fig:vrteffVsRDV} Vertex reconstruction efficiency as a function of radial position $R$. The efficiency is defined as the probability for a true LLP decay to be matched with a reconstructed DV fulfilling the vertex preselection criteria. In \protect\subref{fig:vertexingeffvsrlrt} the efficiencies with and without the special LRT processing are shown for one benchmark signal, while \protect\subref{fig:vertexingeffvsr} shows two $R$-hadron signal samples with different gluino--neutralino mass differences when using LRT processing.}
\end{center}
\end{figure}

Figure~\ref{fig:vrteffVsRDV} shows the DV reconstruction efficiency, defined as the probability for a true LLP decay to be matched with a reconstructed DV fulfilling the vertex preselection criteria (described in \Sect{\ref{sec:eventandvertexselections}}) as a function of $R$. The improvement with respect to standard tracking at large radii is shown in Figure~\ref{fig:vertexingeffvsrlrt}, while Figure~\ref{fig:vertexingeffvsr} shows how the efficiency of the LRT-based DV reconstruction depends on the mass difference $\Delta m = m_{\gluino} - m_{\neutralino}$. With larger mass difference, more and higher-$\pT$ particles are produced in the gluino decay, which increases the reconstruction efficiency of the DV. 

\subsection{Material-dominated regions and the effect of disabled detector modules}
\label{sec:vertexvetoes}

\begin{figure}[!hbtp]
\begin{center}
\subfloat[]{
	\label{fig:dvxyvetoed}
	\includegraphics[width=0.48\textwidth]{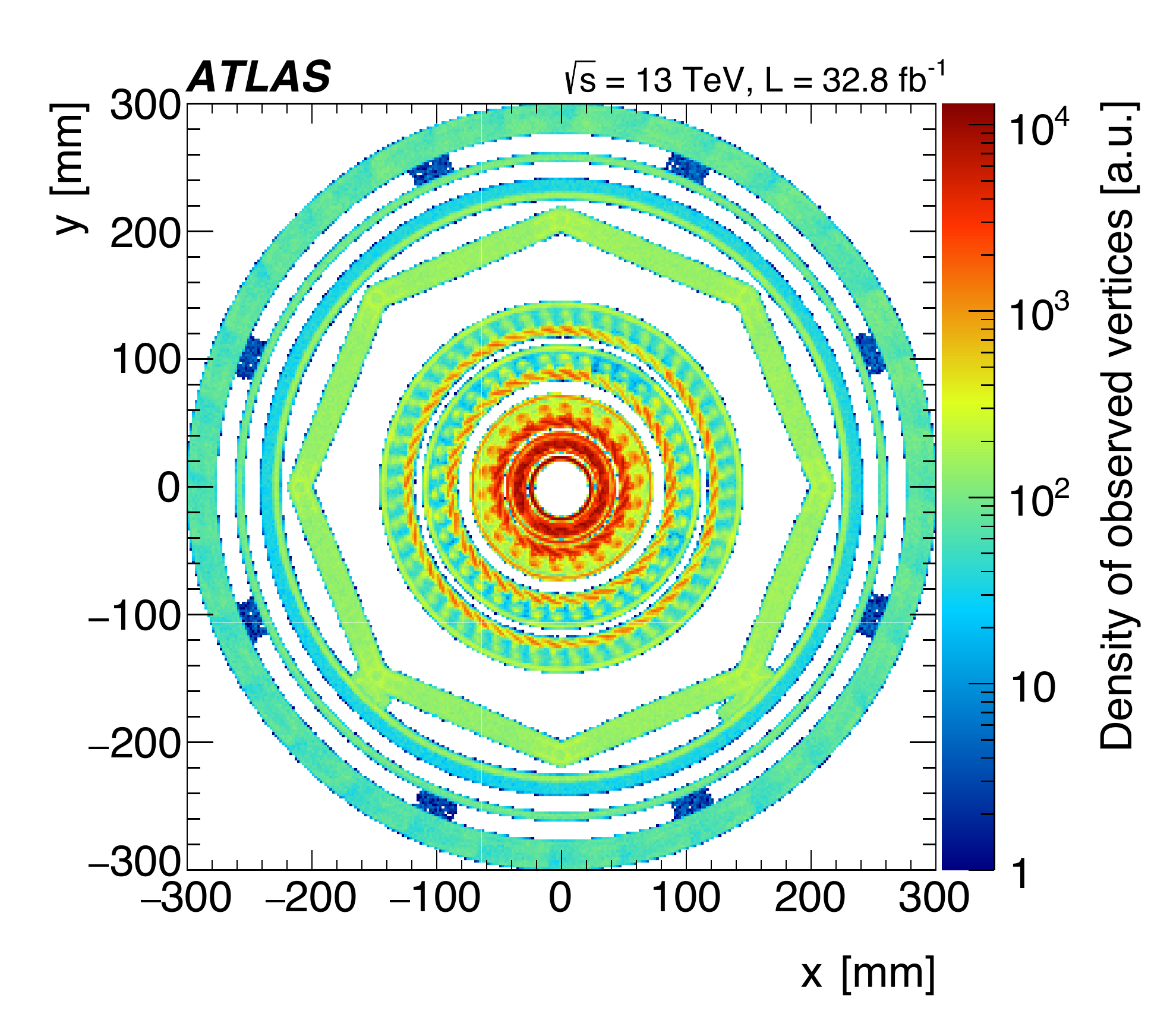}
}
\subfloat[]{
	\label{fig:dvrzvetoed}
	\includegraphics[width=0.48\textwidth]{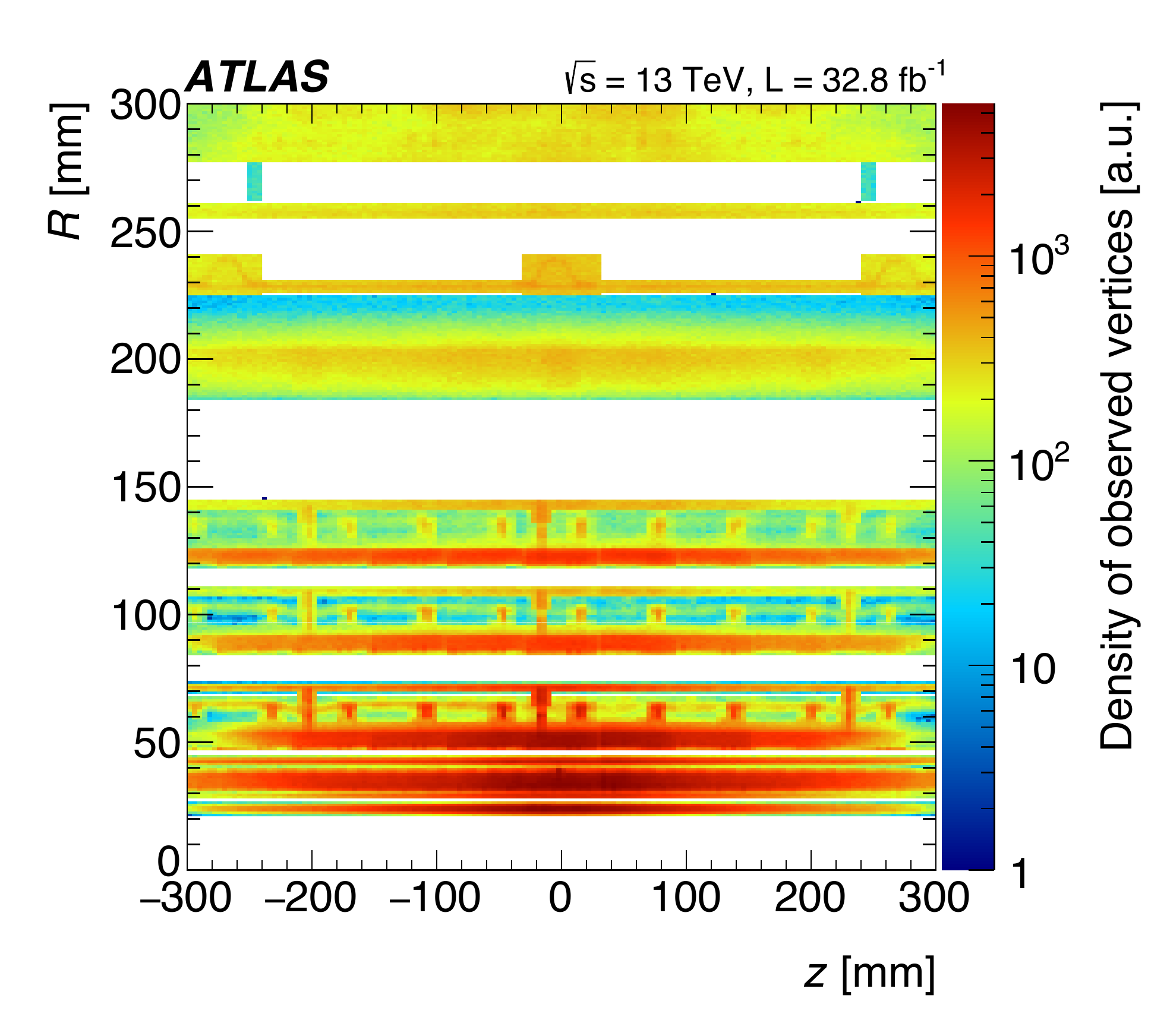}
}
\caption{\label{fig:materialMapValid} Two-dimensional maps of the observed vertex density in regions vetoed by the material map, projected in the \protect\subref{fig:dvxyvetoed} $x$--$y$ plane and \protect\subref{fig:dvrzvetoed} $z$--$R$ plane. The color scale is in arbitrary units (a.u.).
}
\end{center}
\end{figure}

An important background in any search for displaced vertices comes from hadronic interactions in material-rich regions of the detector~\cite{Aaboud:2016poq,PERF-2015-07}. In order to suppress this background, a map defining regions with known material is constructed by studying the positions of DVs in $\sqrt{s}=13$~\TeV\ minimum-bias data. The map is used to reject vertices within the material regions. In these studies, the vertices from the long-lived SM hadrons \Kshort and $\Lambda^0$ are vetoed by discarding vertices that match their expected track multiplicities and reconstructed masses. The application of the map-based veto significantly reduces the contribution from hadronic interactions at the cost of discarding approximately 42\% of the fiducial volume. The material map is visualized in Figure~\ref{fig:materialMapValid}, in which the locations of the observed vertices failing this veto are projected onto the $x$--$y$ and $R$--$z$ planes.

In addition to the material veto map, a veto is applied to reject vertices in regions sensitive to the effect of disabled pixel modules. This requirement discards 2.3\% of the total fiducial volume.

\subsection{Event and vertex selections}
\label{sec:eventandvertexselections}

All events used in this analysis must satisfy the following selection requirements. Firstly, the data was passed through a filter during prompt reconstruction and was made available in a raw data format in order to facilitate the special processing with dedicated track and DV reconstruction required by this analysis. This filtering included passing an $\MET$, multijet, or single-lepton trigger. For the $\MET$-triggered events used in the signal region (SR) of this search, an additional requirement is imposed on \emph{hadronic \MET}, a quantity similar to $\MET$ but with all clusters of energy deposited in the calorimeter calibrated as if they come from hadrons. The filtering of the first 75\% of the data set also required the presence of one trackless\footnote{A jet is considered trackless if $\sum \pT^{\mathrm{track}} < 5$~\GeV, where the sum is taken over all tracks reconstructed in the first reconstruction pass matched to both the PV and the jet.} jet with $\pT > 70$~\GeV\ or two trackless jets with $\pT > 25$~\GeV, and hadronic $\MET > 130$~\GeV. For the last 25\% of the data set, the trackless jet requirement was removed and hadronic $\MET > 180$~\GeV\ was required instead. 
This change was made in order to improve sensitivity for low-$\Delta m$ signal scenarios~\cite{Nagata:2015pra,Nagata:2015hha,Nagata:2017gci}, which are unlikely to give rise to jets with high $\pT$ from the displaced decays. 
The MC events used in this analysis were processed separately in two subsamples with sizes proportional to the integrated luminosities of the two subsamples.

Additional detector-level quality requirements are applied, vetoing events that are affected by calorimeter noise, data corruption, or other effects occurring at the time the data were recorded. Events are required to have at least one PV. To mitigate the contamination of high-$\MET$ events from non-collision background (NCB) processes such as beam halo, additional quality requirements are placed on the leading jet in each event. These requirements use the longitudinal calorimeter-sampling profile of these jets to select for high-$\pT$ hadronic activity originating within the detector volume and reduce NCB contributions to at most $10\%$ early in the event selection. Together with the requirement that such events contain a DV candidate, these criteria are called the \emph{event preselection} and, along with additional DV requirements, are used in the construction of the control region (CR).

To further improve signal sensitivity, the \emph{full event selection} criteria that are used in the construction of the SR require that the event be recorded by an $\MET$ trigger and satisfy $\MET>250$~\GeV. This last requirement ensures that the events are in the plateau of the efficiency turn-on curve for both the $\MET$ trigger and the requirement on the hadronic $\MET$ described above.

The DV candidates are required to satisfy the following conditions, referred to as the \emph{vertex preselection}:
\begin{enumerate}
	\item The vertex position must be within the fiducial volume $R<300$~mm and $|z|<300$~mm.
	\item The vertex must be separated by at least $4$~mm in the transverse plane from all reconstructed PVs.
	\item The vertex must not be in a region that is material-rich or affected by disabled detector modules, as described in \Sect{\ref{sec:vertexvetoes}}.
	\item The vertex fit must have $\chi^2 / \ndof < 5$.
\end{enumerate}

These vertex preselection criteria ensure high-quality measurements of the DV properties and reduce the number of vertices from instrumental effects. Vertices satisfying these criteria are used in the background estimation. For the \emph{final vertex selection} used in the SR of this search, vertices are required to have at least five associated tracks and a reconstructed invariant mass $\mDV>10$~\GeV. These stricter requirements allow the use of vertices with lower mass and 3--4 tracks for building and validating background estimates, and give a low-background search with good signal sensitivities for a large part of the parameter space for the models of interest.

\begin{figure}[!hbtp]
\begin{center}
\subfloat[Fixed $\tau = 1$~ns]{
	\label{fig:signalaccefffixedtau1ns}
	\includegraphics[width=0.5\textwidth]{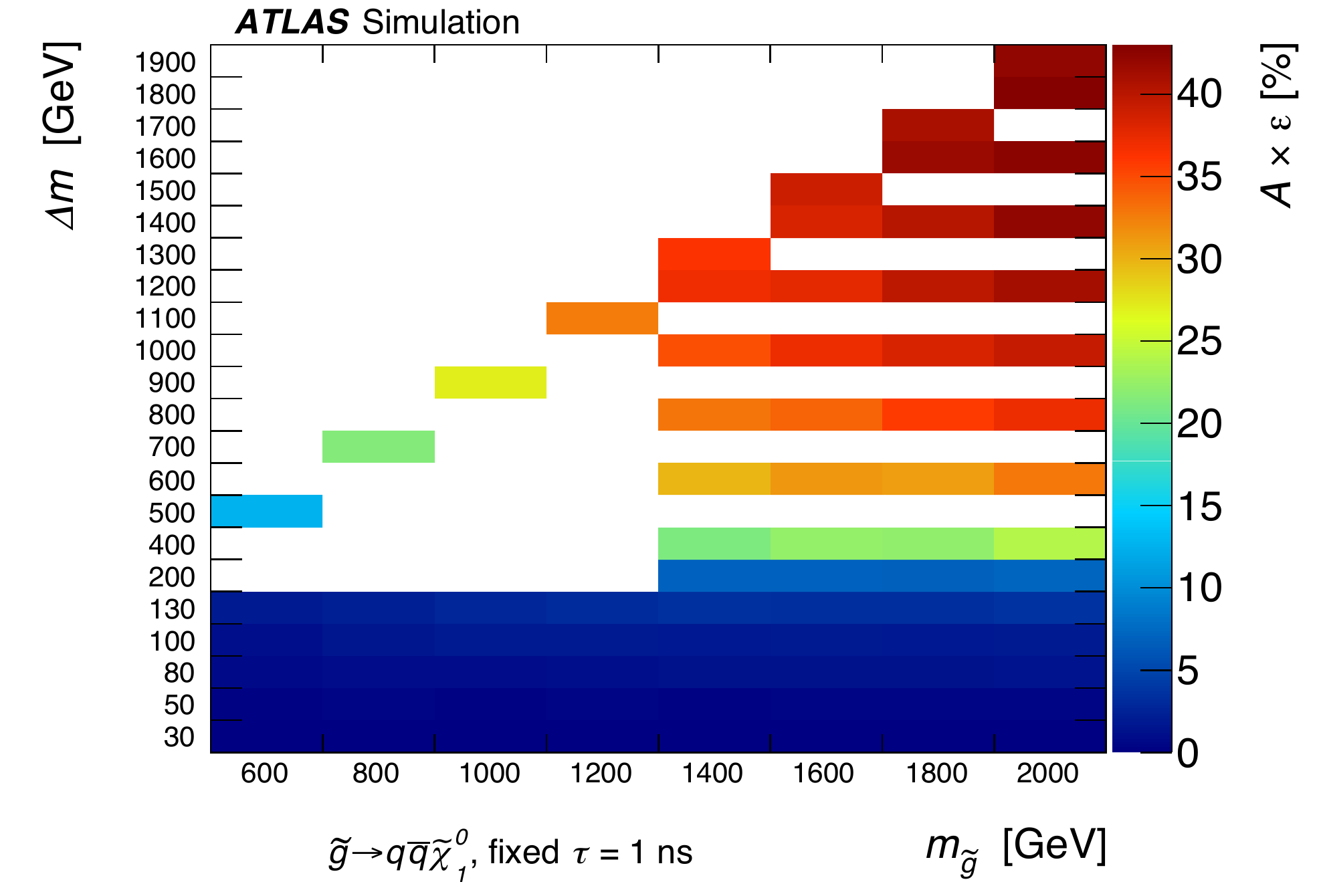}
}
\subfloat[Fixed $\Delta m = 100$~\GeV]{
	\label{fig:signalaccefffixeddm100}
	\includegraphics[width=0.5\textwidth]{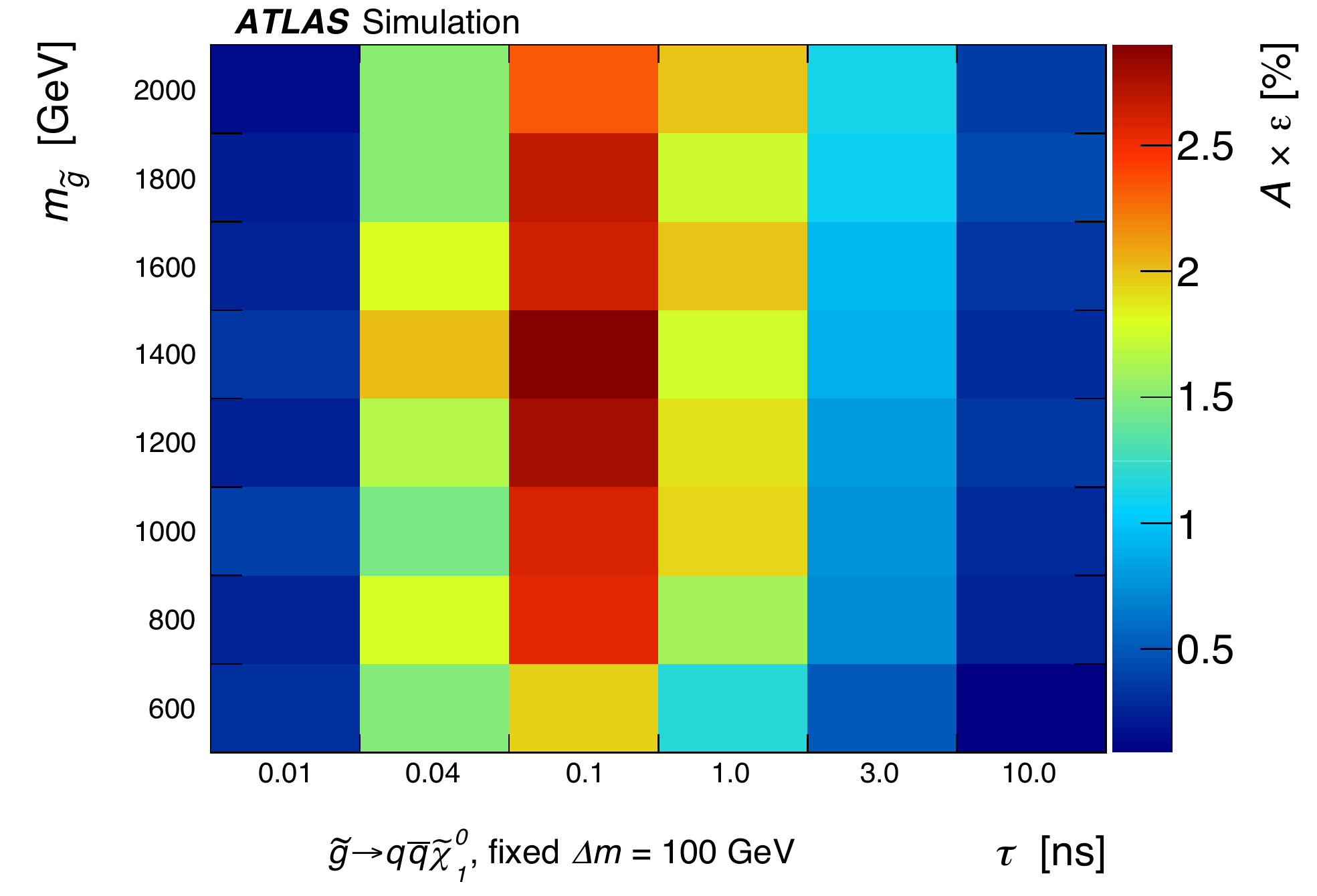}
}
\caption{\label{fig:signalacceff} Fractions of selected events for several signal MC samples, illustrating how $\mathcal{A}\times\varepsilon$ varies with the model parameters. In \protect\subref{fig:signalaccefffixedtau1ns} the gluino lifetime $\tau$ is fixed to 1~ns, and in \protect\subref{fig:signalaccefffixeddm100} the mass difference $\Deltam$ is fixed at 100~\GeV.}
\end{center}
\end{figure}

Figure~\ref{fig:signalacceff} shows the acceptance times efficiency ($\mathcal{A} \times \varepsilon$) of the SR, for several benchmark signal models. In Figure~\ref{fig:signalaccefffixedtau1ns}, the $\mathcal{A} \times \varepsilon$ is shown for models with different gluino and neutralino masses but fixed lifetime of 1~ns. The $\mathcal{A} \times \varepsilon$ depends strongly on the gluino--neutralino mass difference, which is directly proportional to the visible DV mass. For models with $m_{\gluino} > 1.5$~\TeV\ and $\Delta m > 1$~\TeV, the search presented here attains an acceptance times efficiency of as much as 40\%. For models with $\Delta m \lesssim 100$~\GeV\ the $\mathcal{A} \times \varepsilon$ is 5\% or lower. In Figure~\ref{fig:signalaccefffixeddm100}, $\Delta m$ is fixed at $100$~\GeV\ while the lifetime $\tau$ is varied within $0.01$~ns $ < \tau < ~10$~ns. The $\mathcal{A} \times \varepsilon$ is highest for lifetimes around 0.1~ns (corresponding to decay lengths of $\mathcal{O}(10)$~mm). Signal models with low $\Delta m$ are less likely to pass both the event- and vertex-level requirements, due to lower intrinsic $\MET$ and smaller visible DV mass.

\section{Background processes and their estimated yields}
\label{sec:backgrounds}

Given the requirements on the mass ($\mDV > 10$~\GeV) and track multiplicity ($\ntracks \geq 5$) placed on the DV candidates in the SR, there is no irreducible background from SM processes. The entirety of the background expected for this search is instrumental in origin. Three sources of such backgrounds are considered in the analysis. Hadronic interactions can give rise to DVs far from the interaction point, especially where there is material in the detector, support structures, and services. Decays of short-lived SM particles can occur close to each other and be combined into high-mass vertices with large track multiplicities, in particular in the regions closest to the beams. Finally, low-mass vertices from decays of SM particles or hadronic interactions can be promoted to higher mass if accidentally crossed by an unrelated track at a large angle. Each source of background is estimated with a dedicated method, and is separately evaluated in $12$ radial detector regions\footnote{The boundaries for these regions are at $R=22, 25, 29, 38, 46, 73, 84, 111, 120, 145, 180,$ and $300$~mm.} divided approximately by material structures in the ID volume within the fiducial region.

To retain a large number of DVs, the estimates below are performed on events satisfying the event preselection criteria. To obtain a final estimate for the SR, an additional \emph{event selection transfer factor}  $\epsilon_{\textsc{sr}}=(5.1\pm2.5)\times 10^{-3}$ is applied. This factor is determined by measuring the efficiency of the full event selection with respect to the preselection. The events used for calculating $\epsilon_{\textsc{sr}}$ are required to have a DV candidate satisfying the vertex preselection. This method relies on the assumption that the mass and track multiplicity distributions of the DVs do not depend on the quantities used in the event selection, which was demonstrated in data to hold within uncertainties. An additional factor $\kappa$ is applied to account for the potential effect of obtaining multiple DVs per event but is found to be consistent with $1.0$ for the region of DV properties probed in this search.

\subsection{Hadronic interactions}
\label{sec:had-int-bgd}
As discussed in \Sect{\ref{sec:recoandevsel}}, the bulk of the hadronic interactions occur in detector regions with dense material, and these are rejected using the material map. However, residual hadronic interactions may survive the selections, either due to imperfections in the material map or from interactions with gas molecules in regions without solid material. The low-mass region of the $\mDV$ distribution is dominated by hadronic interactions. Therefore, to estimate this background in the SR, the $\mDV$ distribution in the region $\mDV<10$~\GeV\ is fit to an exponential distribution and extrapolated to the SR with $\mDV>10$~\GeV. The assumptions made by this method and the related uncertainties are discussed in \Sect{\ref{sec:uncertainties}}.

\subsection{Merged vertices}
\label{sec:merged-vertices-bgd}

\begin{figure}[!bp]
\begin{center}
\subfloat[Distances between pairs of two-track vertices]{
	\label{fig:mergedtwoplustwo}
	\includegraphics[width=0.47\textwidth]{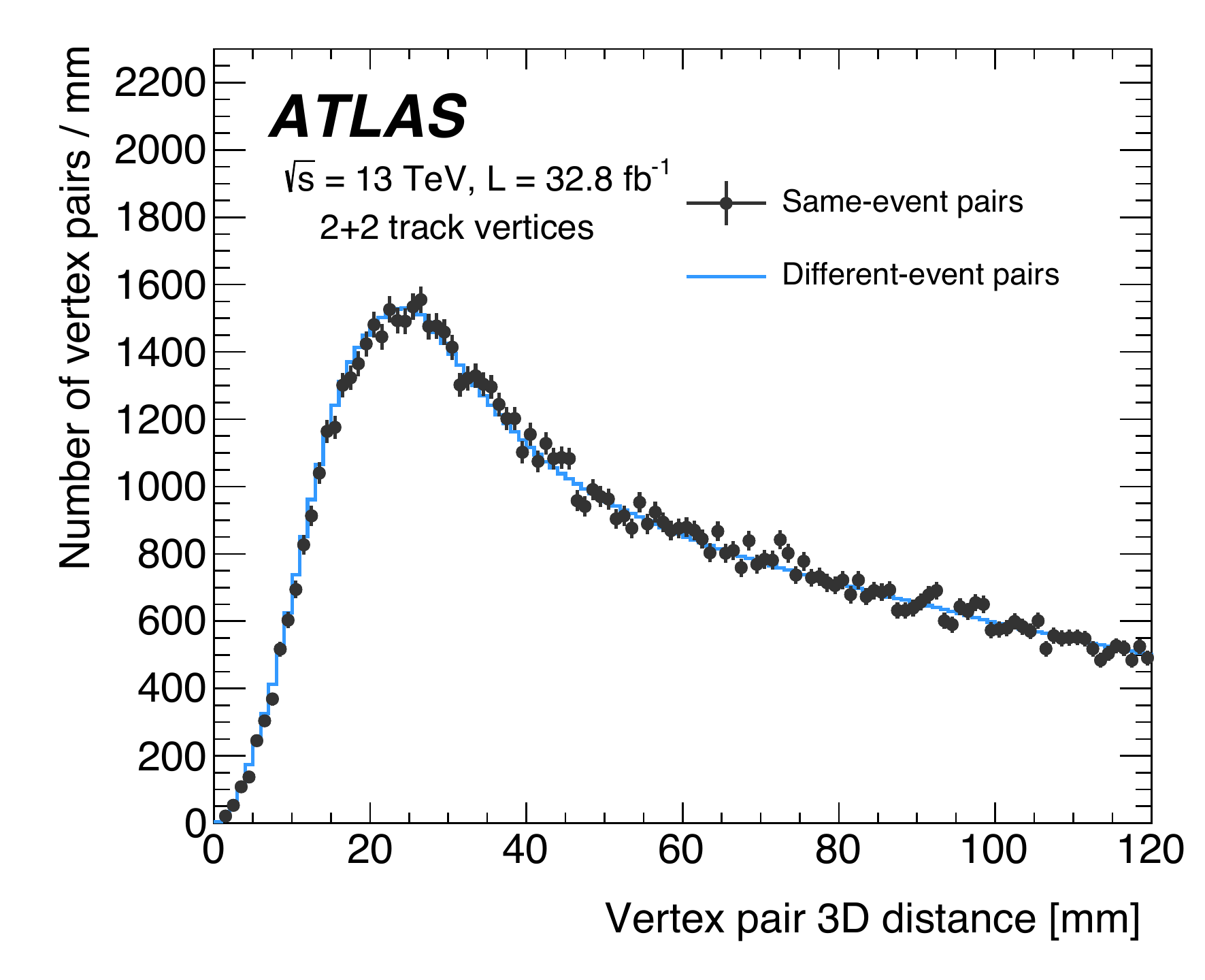}
}
\subfloat[Close-up of the small-distance part of the (2+3) pairs]{
	\label{fig:mergedtwoplusthree_zoomed}
	\includegraphics[width=0.47\textwidth]{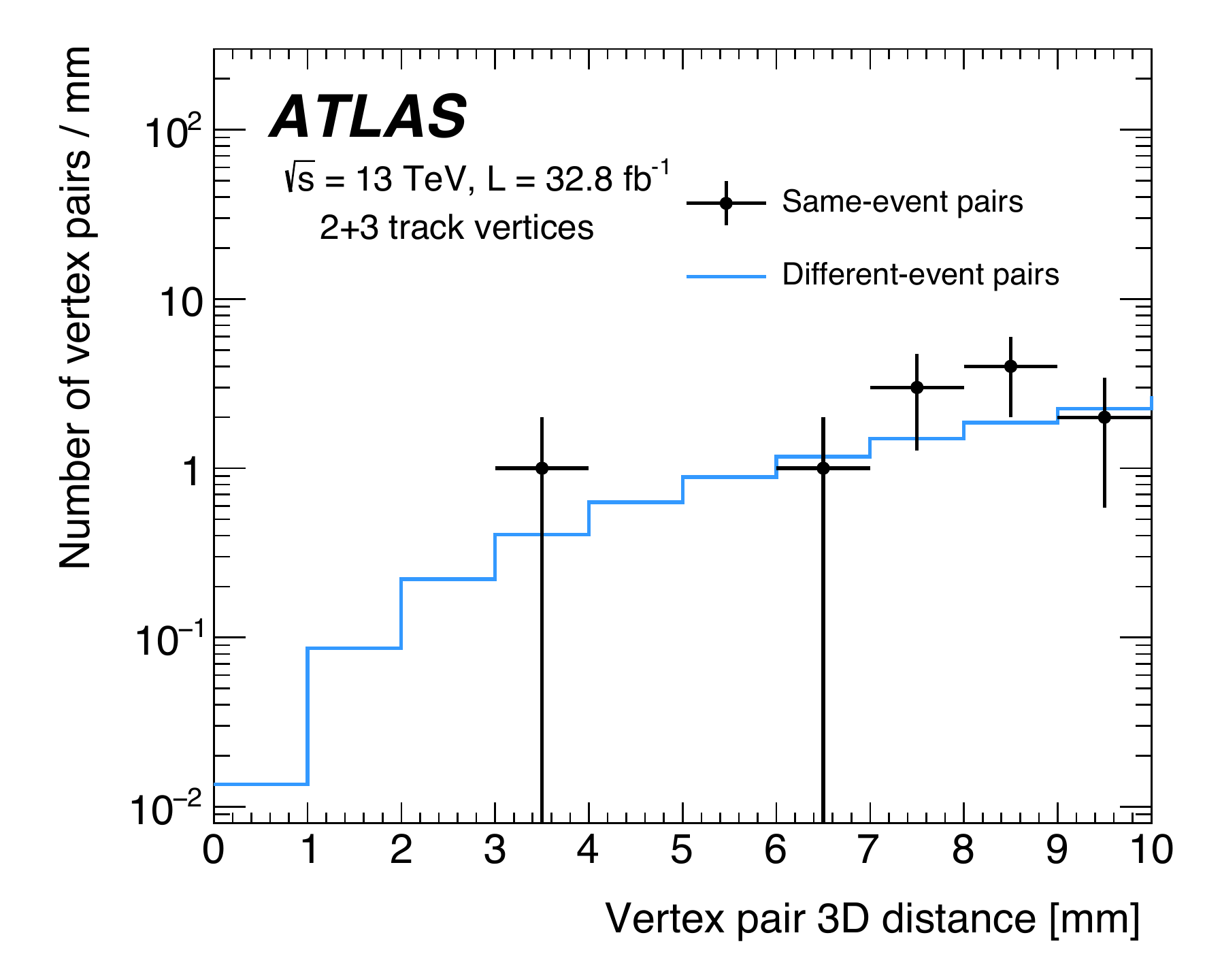}
}
\caption{Distributions of inter-vertex distances in reweighted pairs of vertices passing the vertex preselection in events passing the event preselection. The same-event (black markers) and different-event (blue histogram) samples are shown for \protect\subref{fig:mergedtwoplustwo} pairs of two-track vertices, and \protect\subref{fig:mergedtwoplusthree_zoomed} the small-distance part of the (2+3)-pair combinations. The model yield for inter-vertex distance lower than 1 mm gives the prediction for the vertices in the high-mass region resulting from merging during DV reconstruction. \label{fig:merged_vertices_after_weighting}}
\end{center}
\end{figure}

The high density of vertices at small radii and the last step of the DV reconstruction, where vertices are combined if they are separated by less than 1~mm, could result in the merging of two DVs with low masses and track multiplicities into a single DV with significantly higher mass and track multiplicity. To quantify this contribution, vertices from distinct events are randomly merged. The distribution of the distance $d(\V_1,\V_2)$ between two 2-track or 3-track vertices $\V_1$ and $\V_2$ is studied. To obtain a large sample of reference DV pairs, $d(\V_1, \V_2)$ is measured in a sample in which $\V_1$ and $\V_2$ are taken from different events. This sample is then compared to the sample constructed only from pairs of vertices appearing in the same event. Each of the vertices in these pairs is required to satisfy the DV preselection criteria, and their combined mass is required to be greater than 10~\GeV. The resulting distributions are shown in Figure~\ref{fig:merged_vertices_after_weighting} for (a) pairs of 2-track vertices (2+2) and (b) for the case of a 2-track vertex paired with a 3-track vertex (2+3). To extract an estimate of the number of SR vertices merged during DV reconstruction, the different-event distribution is normalized to the same-event distribution in the region $d(\V_1,\V_2) > 1$~mm, and the estimated contribution from merged vertices is given by the scaled template's integral for $d(\V_1,\V_2) < 1$~mm.

It is found that the $z$ positions of $\V_1$ and $\V_2$ in the same-event sample are correlated, since they are likely to originate from the same hard-scatter primary vertex. Naturally, this effect is absent in the different-event sample. As a result, the distributions of the longitudinal distance between the vertices in the different-event and same-event samples differ by up to 30\% at low values of $d(\V_1,\V_2)$. To correct for this difference between the two samples, the DV pairs in the different-event sample are reweighted to match the distribution of distances in $z$ in the same-event sample before the yield for $d(\V_1,\V_2) < 1$~mm is extracted. After applying the weights, the model distribution of the three-dimensional distance $d(\V_1,\V_2)$ agrees well with that of the same-event sample in the studied range of $d(\V_1,\V_2) < 120$~mm. This reweighting procedure is applied in the distributions shown in Figure~\ref{fig:merged_vertices_after_weighting}.

The background from merged DV pairs with $d(\V_1,\V_2) < 1$~mm and $\ntracks \geq 5$ tracks is estimated from DV pairs where one DV has two tracks and the other has three tracks. This background is found to be orders of magnitude smaller than the accidental-crossing background discussed below. The background from the merging of two 3-track vertices or a 2-track and a 4-track DV is determined to be negligible compared to other sources for higher track multiplicities.

\subsection{Accidental crossing of vertices and tracks}
\label{sec:accidental-crossing-bgd}

The final and dominant source of background in the SR for this search is low-mass vertices crossed by an unrelated track in the event. It is common for such crossings to occur at large angles with respect to the \emph{distance vector} that points from the PV to the DV. This significantly increases the mass of the DV. In order to estimate the contribution from this effect, ($n$+1)-track vertices are constructed by adding a \emph{pseudo-track} to $n$-track vertices from the data. The pseudo-track is given track parameters drawn randomly from track templates, extracted separately for each radial detector region. These templates are constructed using all tracks associated with DV candidates satisfying $\ntracks \geq 3$ and $\mDV>3$~\GeV\ found in events passing the event preselection. The templates contain the track $\pT$, $\eta$, and relative azimuthal angle $\Delta\phi$ with respect to the distance vector. In order to model the effect of high-angle crossings, pseudo-tracks drawn from the templates are required to be at an angle larger than $\sqrt{(\Delta \eta)^2+(\Delta \phi)^2}=1$ with respect to the distance vector.

\begin{figure}[!hbtp]
\centering
\subfloat[][Before the beam pipe \\($R < 22$~mm)]{
	\label{fig:dv_mass_fitter_3Trk_Region0}
	\includegraphics[width=0.48\textwidth]{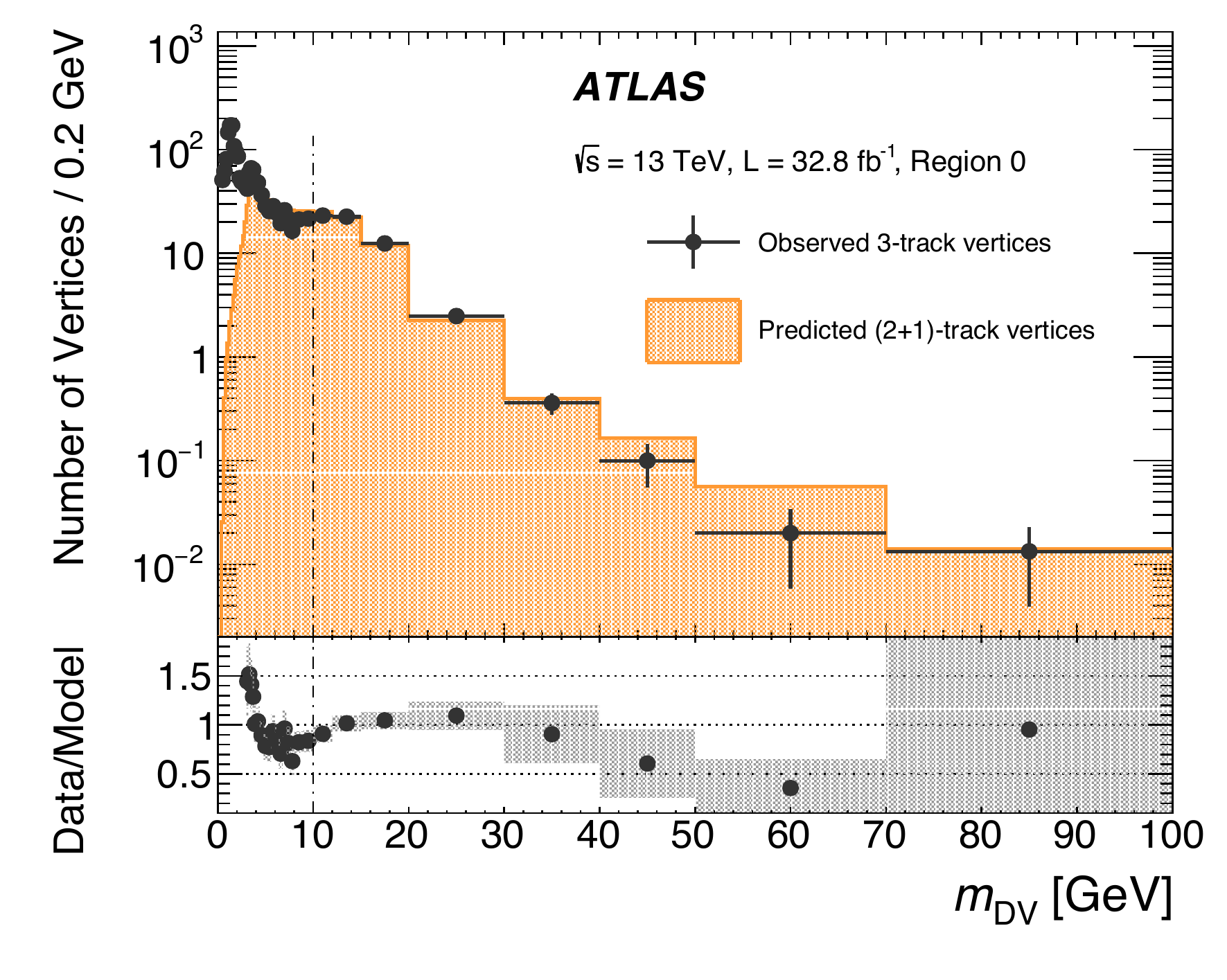}
}
\subfloat[][Before pixel layer 1 \\($73$~mm $< R < 84$~mm)]{
	\label{fig:dv_mass_fitter_3Trk_Region6}
	\includegraphics[width=0.48\textwidth]{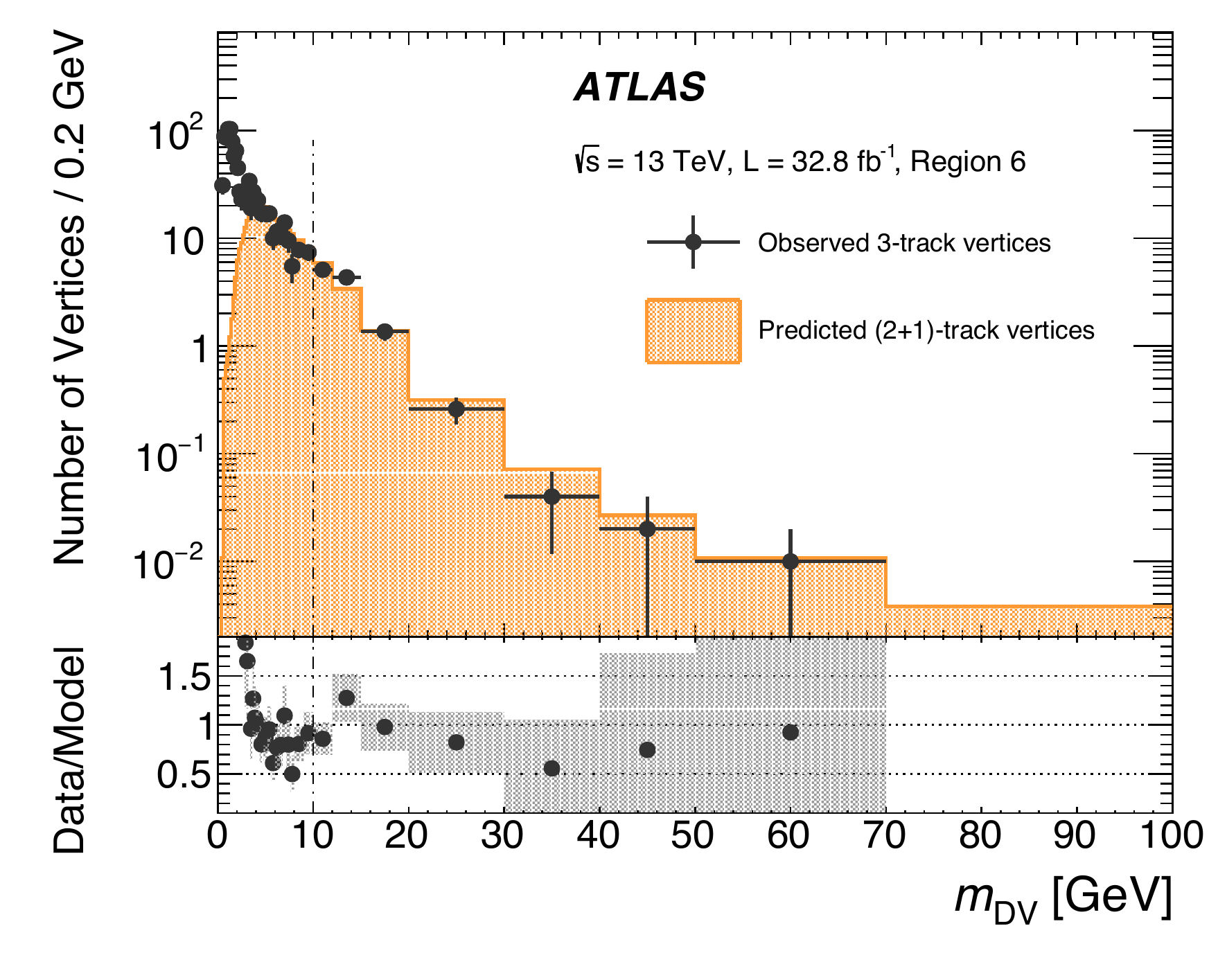}
}
\caption{\label{fig:dv_mass_fitter_3TrkSelectedRegions} Distributions of $\mDV$ for 3-track vertices in the CR data for two radial regions, along with the normalized predictions from the track-association method. The spectra from the model are normalized to the data in the $\mDV>10$~\GeV\ region, and the scaling needed is extracted and used as the crossing factors used to calculate the predictions for higher track multiplicities. The error bars and the gray bands in the bottom ratio distributions represent the statistical uncertainties. The region below 10~\GeV\ is not expected to be described by the accidental-crossing model.}
\end{figure}

To normalize the prediction from the model constructed by this method, the probability of an accidentally crossing track to become associated with the DV is extracted by comparing the sample of $3$-track vertices seen in the data to the (2+1)-track vertices from the model in the $\mDV > 10$~\GeV\ region. This probability is referred to as the \emph{crossing factor} and is extracted separately for each radial detector region. Figure~\ref{fig:dv_mass_fitter_3TrkSelectedRegions} shows the resulting (2+1)-track predictions from the model along with the 3-track vertices for two selected radial regions. The observed differences in shape between the model and the data are used in Section~\ref{sec:uncertainties} to assess an uncertainty in the background estimates from the model. These crossing factors are used to project from an $n$-track CR to an ($n$+1)-track region for events passing the event preselection.

\subsection{Validation of background estimation techniques}

To ensure that the methods described above reliably model the backgrounds, two validation regions are constructed and used to test their predictions. The two regions are designed to be free of significant contamination from any signal considered in this analysis. In a low-$\MET$ validation region, denoted \textsc{vrlm}, the performance of these methods for vertices with exactly four tracks is studied as an intermediate point between the $3$-track CR and the $\geq 5$-track SR. The \textsc{vrlm} event selection requires $\MET < 150$~\GeV\ and that the minimum azimuthal angle between the $\MET$ vector and all reconstructed jets, $\mindphi$, is less than $0.75$. These requirements sufficiently reduce the contribution from the considered signal processes that are not excluded by previous searches~\cite{Aad:2015rba}. The background estimate extracted from the CR is scaled to account for the efficiency $\epsilon_{\textsc{vrlm}}$ of the $\MET$ and $\mindphi$ requirements to predict the background in \textsc{vrlm}. Since studies in data show that the $\mDV$ and $\ntracks$ distributions are independent of these event-level quantities, $\epsilon_{\textsc{vrlm}}$ is extracted in a sample with $3$-track vertices and applied to the 4-track prediction. It is found to be $\epsilon_{\textsc{vrlm}}=(56\pm6)\%$.

Additional validation of the background estimation methods is done in a material-enriched validation region, \textsc{vrm}. Here, the material veto is inverted and vertices satisfying the other vertex preselection criteria are studied. Due to the abundance of hadronic interactions in this region, it contains many more vertices than \textsc{vrlm}. Since accidental track crossings also happen to vertices from hadronic interactions, this region can be used to validate the accidental-crossing background estimation method. An independent set of crossing factors are derived and applied in this validation region, and their values are found to be similar to those extracted in the samples where the material-rich regions are vetoed.

In both \textsc{vrlm} and \textsc{vrm}, the yields predicted by the background estimation methods are shown in Table~\ref{tab:BkgVertices}.

\subsection{Final expected yields}
\label{sec:finalbgestimates}

The predicted background yields in the various selections are listed in \Tab{\ref{tab:BkgVertices}}. The yields are shown separately for each of the estimation methods along with the total for each region. Also shown is the final expected yield in the SR after the application of the scaling factors described above. The total SR prediction from the sum of all background sources is $0.02^{+0.02}_{-0.01}$ events, where the total uncertainty includes both the statistical and systematic uncertainties.

\begin{table}
    \begin{center}
        \caption{The number of estimated background vertices with mass $\mDV > 10$~\GeV\ for the DV selections used in the control and validation regions are shown. The ($n+1$)-track contributions are estimated using the accidental-crossing factor method (\Sect{\ref{sec:accidental-crossing-bgd}}), the ($2+i$)-track contribution is obtained from merged vertices (\Sect{\ref{sec:merged-vertices-bgd}}), and the pure $n$-track contribution is evaluated using the hadronic interactions (\Sect{\ref{sec:had-int-bgd}}). Also shown are the estimated background event yields in the preselection region with at least five tracks. The predicted background event yield in the signal region appears in the bottom row and includes the transfer factors shown. When two uncertainties are shown, the first is statistical while the second is systematic. When one number is given, it represents the combined uncertainty.
        }\label{tab:BkgVertices} \small
        \begin{tabular}{l l l r@{\hskip6pt}c@{\hskip6pt}r@{\hskip6pt}c@{\hskip6pt}r} \hline \hline
          Selection   &  Subregion  &   Category  & \multicolumn{5}{r}{Yield}   \\ \hline
          &  &  & \\
          \parbox{4cm}{\emph{Event preselection}\\ $n_{\text{trk}}=3,\ \mDV>10$~\GeV} &  & Measured total & \multicolumn{5}{r}{$3093$} \\
          &  &  & \\
          \hline
          &  &  & \\
          \multirow{8}{*}{\parbox{4cm}{\emph{Event preselection}\\ $n_{\text{trk}}=4,\ \mDV>10$~\GeV}}
          & \multirow{1}{*}{\textsc{vrlm}} & $(3+1)$-track		& $12.6$	& $\pm$	& $0.3$	& $\pm$	& $1.1$		\\
          &                                		& $(2+2)$-track		& $3.6$	& $\pm$ 	& & $3.6$	&  	\\
          &                                		& Pure $4$-track	& $0.3$	& 		& & $^{+0.9}_{-0.3}$	&  	\\
          &                                		& \it Subtotal		& $16$	& $\pm$ 	& & $4$	&  	\\
          &                                		& \textit{Total}\quad (after scaling by $\epsilon_{\textsc{vrlm}}$)	& $9$ & $\pm$ & & $2$ & \\[2ex]

          & \multirow{1}{*}{\textsc{vrm}} 	& $(3+1)$-track  	& $137$	& $\pm$	& $3$	& $\pm$	& $30$		\\
          &                               		& Pure $4$-track 	& $16$	& 		& & $^{+47}_{-16}$	&	\\
          &                               		& \it Total      		& $150$	& 		& & $^{+60}_{-30}$	&	\\[2ex]
          \hline \\[2ex]
        	  \multirow{12}{*}{\parbox{4cm}{\emph{Event preselection}\\ $n_{\text{trk}}\ge5,\ \mDV>10$~\GeV}}
          & \multirow{1}{*}{$5$-tracks} 	& $(4+1)$-track		& $1.30$	& $\pm$	& $0.07$	& $\pm$	& $0.12$		\\
          &                             			& $(2+3)$-track		& $0.01$	& $\pm$	& & $0.01$	&	\\
          &                             			& Pure $5$-track	& $0.9$ 	& 	& & $^{+2.8}_{-0.9}$		&	\\
          &                             			& \it Total			& $2.2$	&	    & & $^{+2.8}_{-0.9}$		&	\\[2ex]

          & \multirow{1}{*}{$6$-tracks} 	& $(5+1)$-track		& $0.37$ 	& $\pm$	& $0.03$ 	& $\pm$ 	& $0.04$		\\
          &                             			& Pure $6$-track	& $0.2$	&	& & $^{+0.6}_{-0.2}$	&	\\
          &                             			& \it Total			& $0.6$	& 	& & $^{+0.6}_{-0.2}$	&	\\[2ex]

          &  \multirow{1}{*}{$\ge7$-tracks} & $(n+1)$-track	& $0.37$	& $\pm$	& $0.03$	& $\pm$	& $0.04$		\\
          &                                 		& Pure $\geq7$-track& $1$	& 	& & $^{+3}_{-1}$	&	\\
          &                                 		& \it Total			& $1$	&   & & $^{+3}_{-1}$	&	\\[4ex]
          &  \multicolumn{2}{l}{\it Total}					& $4.2$	&	& & $^{+4.1}_{-1.4}$	&	\\[2ex]
          \hline\hline \\[0.5ex]
          \it Full SR selection			& \multicolumn{2}{l}{\textit{Total}\quad (after scaling by $\epsilon_{\textsc{sr}} \times \kappa$) }	& $0.02$ & 	& 	& $^{+0.02}_{-0.01}$ &   \\[2ex]
          \hline\hline
        \end{tabular}
    \end{center}
\end{table}

\section{Uncertainties}
\label{sec:uncertainties}

The estimation of the hadronic interaction background described in \Sect{\ref{sec:had-int-bgd}} relies on the assumption that the mass spectra of such contributions follow an exponential shape. This assumption is tested using interaction vertices in the \geant-based simulations described in \Sect{\ref{sec:datasets}}. Based on studies of the deviations from an exponential shape seen in the simulation, an uncertainty of $-100\%$ and $+300\%$ is applied to the component of the total background from hadronic interactions. The size of this uncertainty is taken as the largest deviation observed in all track multiplicities for vertices with $\mDV>10$~\GeV\ in simulation.

The background in the SR due to merged vertices (\Sect{\ref{sec:merged-vertices-bgd}}) is estimated to be very small with respect to the total background. By comparing the same-event data and different-event model for (2+3)-track DV pairs, the largest statistically significant discrepancy in any bin in the studied range is observed to be $60$\%. To be conservative, the systematic uncertainty for this subdominant background is taken to be $100\%$.

Uncertainties associated with the contribution from low-mass vertices crossed accidentally by an unrelated track (\Sect{\ref{sec:accidental-crossing-bgd}}) are dominated by the uncertainty of the extracted crossing factors. By varying the choice of $\mDV$ threshold used for the normalization of the spectra from the background model by $\pm5$~\GeV\ (with respect to the nominal $10$~\GeV), an uncertainty is extracted. Since the crossing factors are derived and applied separately for each radial detector region, their uncertainties are as well. The size of the resulting uncertainty for the accidentally crossing track contributions is $10$--$20\%$ depending on the radial detector region.

Finally, the event selection transfer factor $\epsilon_{\textsc{sr}}$ and the correction $\kappa$ from event level to vertex level, described in \Sect{\ref{sec:backgrounds}}, also have associated uncertainties. Both of these uncertainties are derived by varying the kinematic requirements for the vertices. Varying the vertex-level requirements used in these calculations results in uncertainties of $50\%$ in $\epsilon_{\textsc{sr}}$ and $16\%$ in $\kappa$. Since these factors are applied to all background contributions to obtain a final SR estimate, these uncertainties propagate directly to the final estimate. 

While the background uncertainties and expectations are derived from data, additional modeling uncertainties that only affect the signal efficiencies are considered and derived by varying parameters used in the simulation and reconstruction. The effect on the signal efficiency due to variations of the amount of simulated pileup is a few percent for high-\Deltam\ samples, and up to $10\%$ for small-\Deltam\ samples. To estimate the size of the uncertainty due to ISR modeling, the size of the reweighting of \PYTHIAV{6} to \MGMCatNLO as described in \Sect{\ref{sec:datasets}} is taken as an additional systematic uncertainty. This effect corresponds to an uncertainty of a few percent in the signal efficiency for high-\Deltam\ models. However, for low-\Deltam\ samples, where the intrinsic \met\ is smaller, the signal acceptance depends heavily on radiation effects. For these models, the uncertainty in the ISR modeling yields an uncertainty of as much as $25\%$ in the acceptance. 

The uncertainty in the signal efficiency due to variations in the track and DV reconstruction efficiency is determined to be $5$--$10\%$ by randomly removing tracks at a rate given by the expected tracking inefficiency. Additional uncertainties involving the reconstructed jet energy scale and resolution, as well as the reconstruction of the $\met$, are evaluated and found to be negligible with respect to the leading uncertainties. No additional uncertainty is considered for the modeling of the production of $R$-hadrons and their interactions with matter. Decays of electrically charged and neutral LLPs are reconstructed as displaced vertices in the ID with similar efficiencies, so this search is less sensitive to the fraction of charged states after hadronization compared to those based on direct-detection signatures. Since the amount of material traversed before a decay in the ID is small, the sensitivity to uncertainties in the per-parton cross section for hadronic interactions is negligible.

\section{Results}
\label{sec:results}

\begin{figure}[!htbp]
\begin{center}
\subfloat[]{  
	\label{fig:massvsntrk_1400_100_1}
	\includegraphics[width=0.47\textwidth]{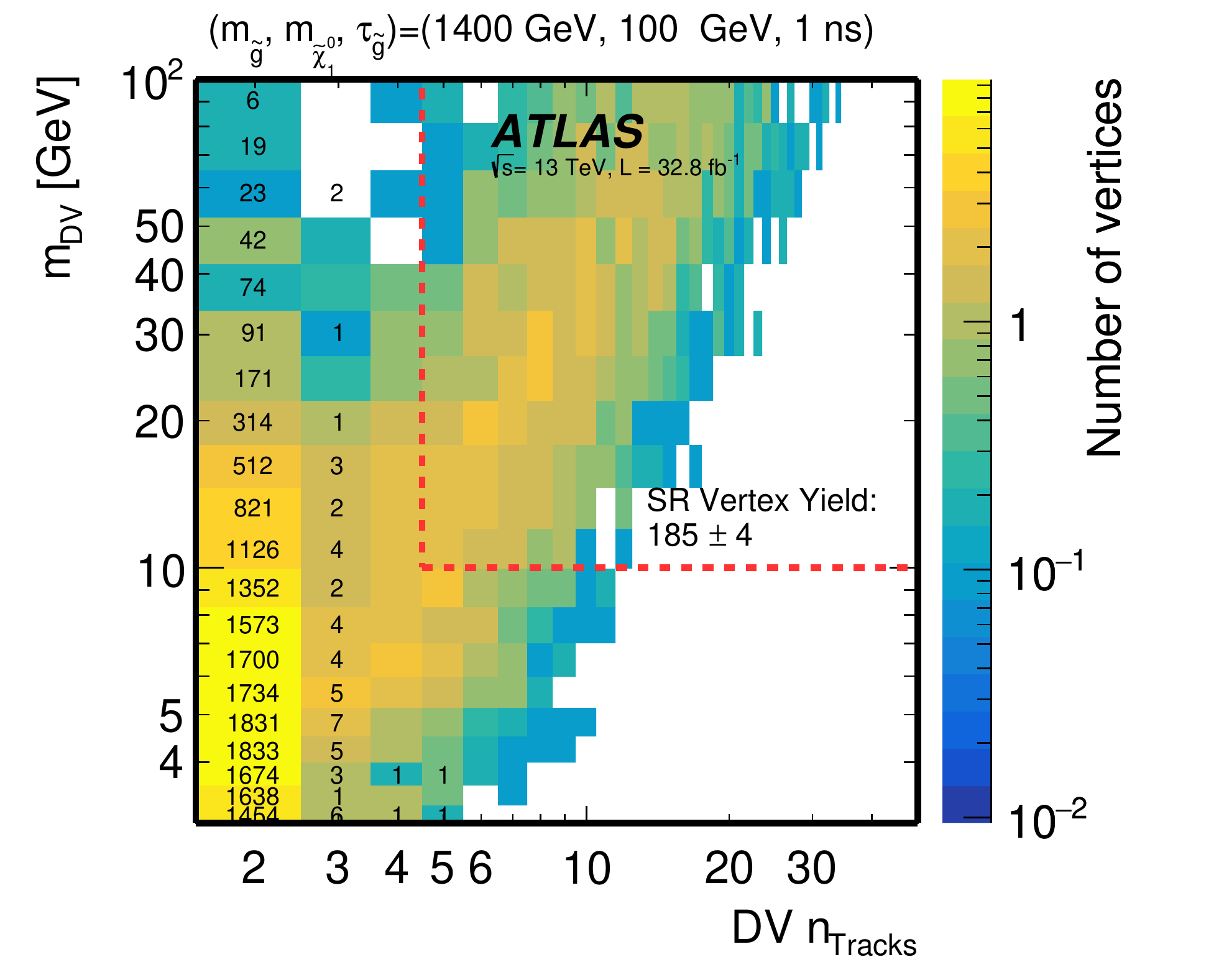}
}
\subfloat[]{ 
	\label{fig:massvsntrk_1400_1320_1}
	\includegraphics[width=0.47\textwidth]{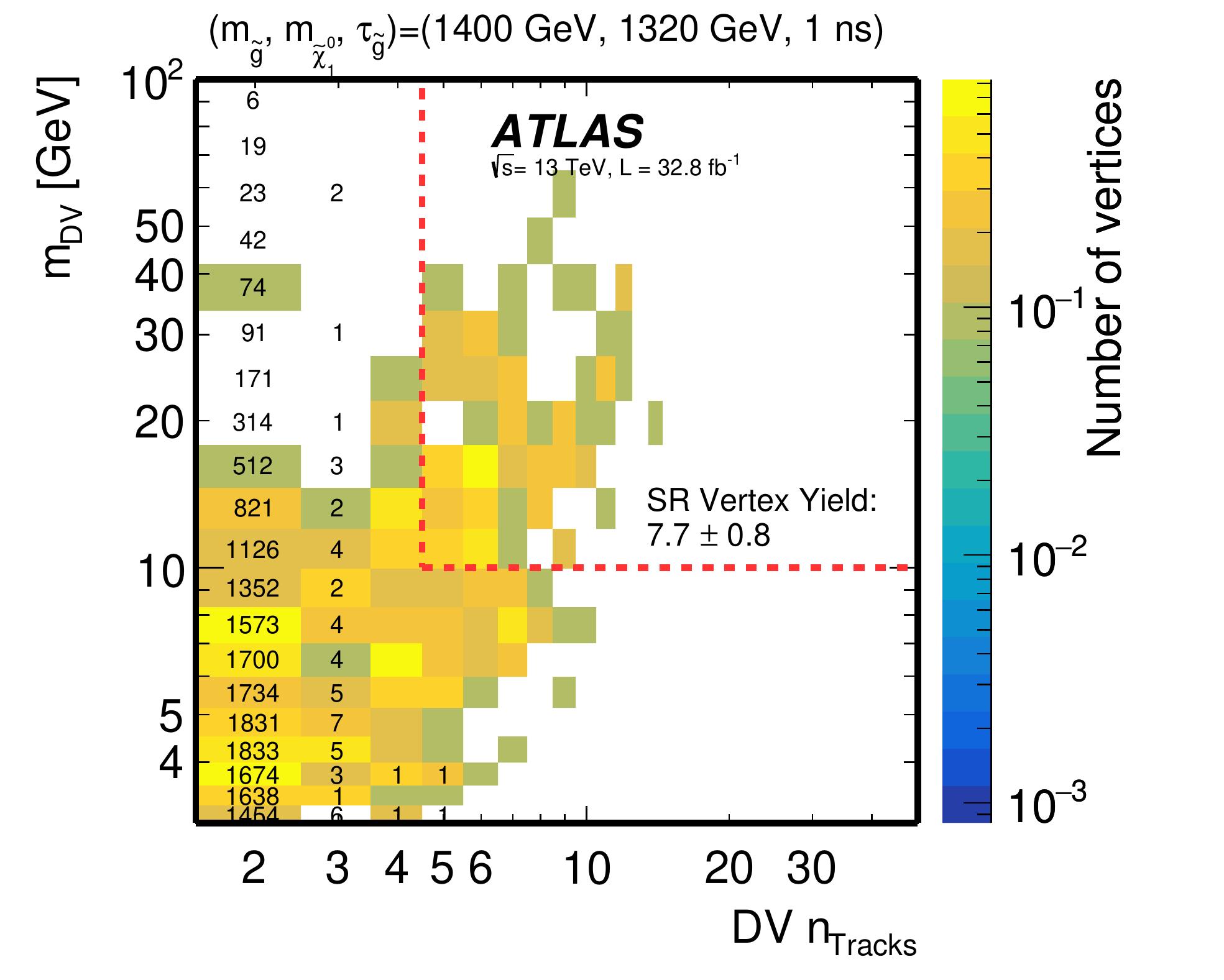}
}
\caption{\label{fig:massvsntrk} Two-dimensional distributions of \mDV\ and track multiplicity are shown for DVs in events that satisfy all signal region event selection criteria. Bin numbers correspond to the observations in data, while the color-representation shows example distributions for two $R$-hadron signals used as benchmark models in this search. The dashed line represents the boundary of the signal region requirements, and the expected signal yield in this region is shown.}
\end{center}
\end{figure}

The final yields for all regions used in this analysis are shown in \Tab{\ref{tab:ResultsTable}}. The observed yields are consistent with the expected background in the validation regions, where \textsc{vrlm} contains $9$ vertices ($9\pm2$ expected) and \textsc{vrm} contains $177$ vertices ($150\pm60$ expected). The two-dimensional distribution of \mDV\ and track multiplicity is shown in \Fig{\ref{fig:massvsntrk}} for events that satisfy the full event-level selection. The final SR yields are highlighted, with $0$ events observed ($0.02^{+0.02}_{-0.01}$ expected).

\begin{table}
    \begin{center}
        \caption{The observed number of vertices for the control and validation regions are shown along with the background expectations. The last row shows the expected and observed signal region event yields.}\label{tab:ResultsTable}
        \small
        \begin{tabular}{l l r@{\hskip3pt}c@{\hskip3pt}r r} \hline \hline
          Selection   &  Subregion  & \multicolumn{3}{r}{Estimated}  & Observed \\
          \hline
			\noalign{\vskip 2mm}
          \parbox{4cm}{\emph{Event preselection}\\ $n_{\text{trk}}=3,\ \mDV>10$~\GeV} & & & & & $3093$\\[0.5ex]
			\noalign{\vskip 2mm}
          \hline
			\noalign{\vskip 2mm}
          \multirow{2}{*}{\parbox{4cm}{\emph{Event preselection}\\ $n_{\text{trk}}=4,\ \mDV>10$~\GeV}}
            &  \textsc{vrlm}	&  $9$	& $\pm$	& $2$	& $9$ \\
            &  \textsc{vrm}	&  $150$	& & $^{+60}_{-30}$	& $177$ \\
			\noalign{\vskip 2mm}
          \hline
  			\noalign{\vskip 2mm}
           \multirow{5}{*}{\parbox{4cm}{\emph{Event preselection}\\ $n_{\text{trk}}\ge5,\ \mDV>10$~\GeV}}
            &  $5$-tracks    	& $2.2$	&	& $^{+2.8}_{-0.9}$	& $1$ \\[0.5ex]
            &  $6$-tracks		& $0.6$	& 	& $^{+0.6}_{-0.2}$	& $1$ \\[0.5ex]
            &  $\ge$7-tracks	& $1$	& 	& $^{+3}_{-1}$	& $3$ \\[2ex]
            &  \it Total		& $4.2$	& 	& $^{+4.1}_{-1.4}$	& $5$ \\
			\noalign{\vskip 2mm}
            \hline\hline
            			\noalign{\vskip 2mm}
            \it Full SR selection &  \it Total	& $0.02$	& 	& $^{+0.02}_{-0.01}$	& $0$ \\
			\noalign{\vskip 2mm}
			 \hline\hline
        \end{tabular}
    \end{center}
\end{table}

\begin{figure}[!htbp]
\begin{center}
\subfloat[Upper limits on production cross section]{
	\label{fig:upperlimit_mlsp_100}
	\includegraphics[width=0.47\textwidth]{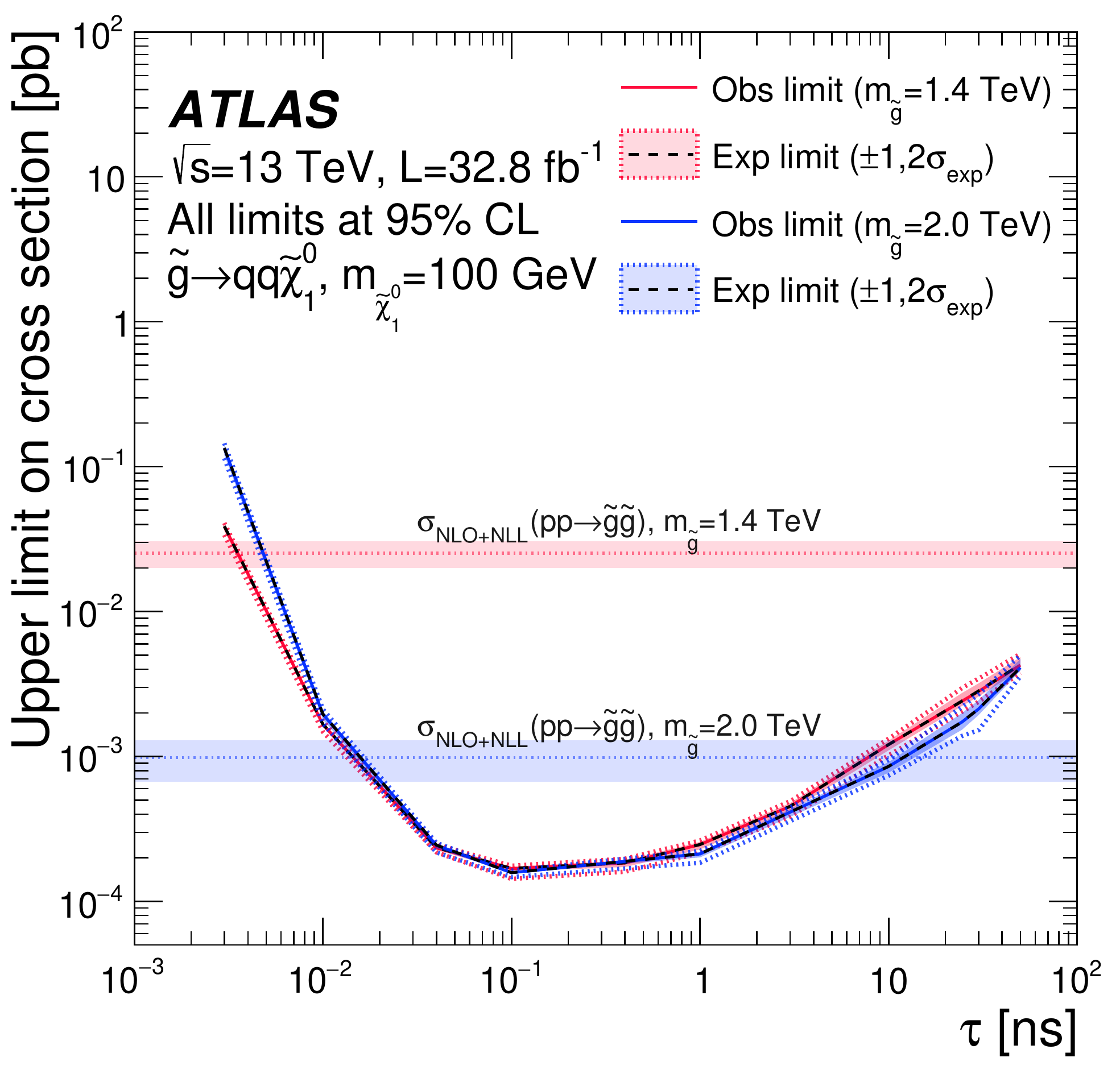}
}
\subfloat[Lower limits on $m_{\gluino}$]{
	\label{fig:mg_vs_ctau_mlsp_100}
	\includegraphics[width=0.47\textwidth]{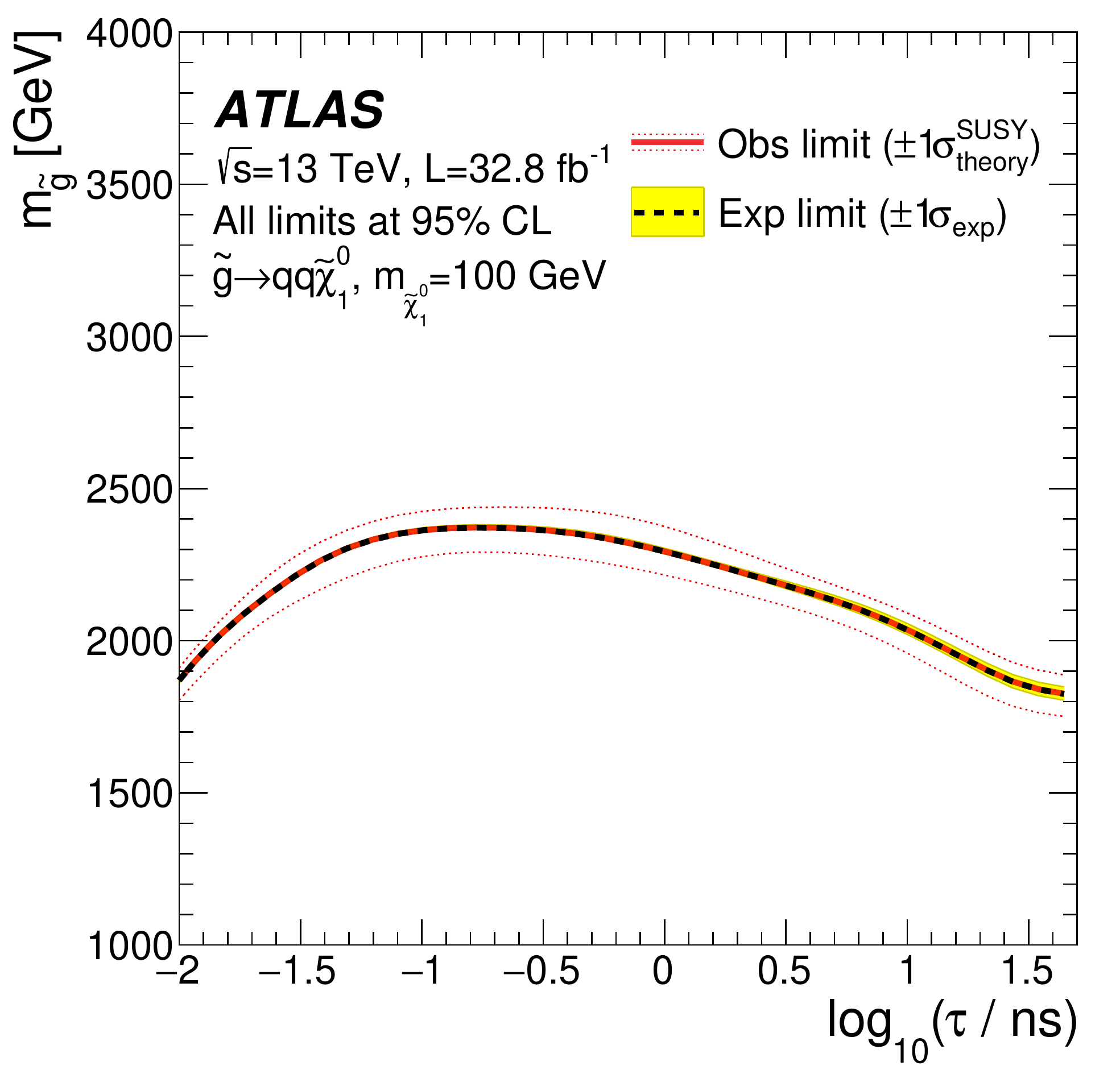}
}
\caption{\label{fig:exclusion_mlsp100} Upper 95\% CL limits on the signal cross section are shown in \protect\subref{fig:upperlimit_mlsp_100} for $m_{\gluino}=1400$~\GeV\ and $m_{\gluino}=2000$~\GeV\ as a function of lifetime $\tau$, for fixed $m_{\tilde{\chi}_{1}^{0}}=100$~\GeV. Horizontal lines denote the $\gluino\gluino$ production cross section for the same values of $m_{\gluino}$, shown with uncertainties given by variations of the renormalization and factorization scale and PDF uncertainties. The lower limit on $m_{\gluino}$ for fixed $m_{\tilde{\chi}_{1}^{0}}=100$~\GeV\ as a function of lifetime $\tau$ is shown in \protect\subref{fig:mg_vs_ctau_mlsp_100}. The nominal expected and observed limit contours coincide due to the signal region yield's high level of agreement with expectation.}
\end{center}
\end{figure}

\begin{figure}[!htbp]
\begin{center}
\subfloat[Upper limits on production cross section]{
	\label{fig:upperlimit_dm_100}
	\includegraphics[width=0.47\textwidth]{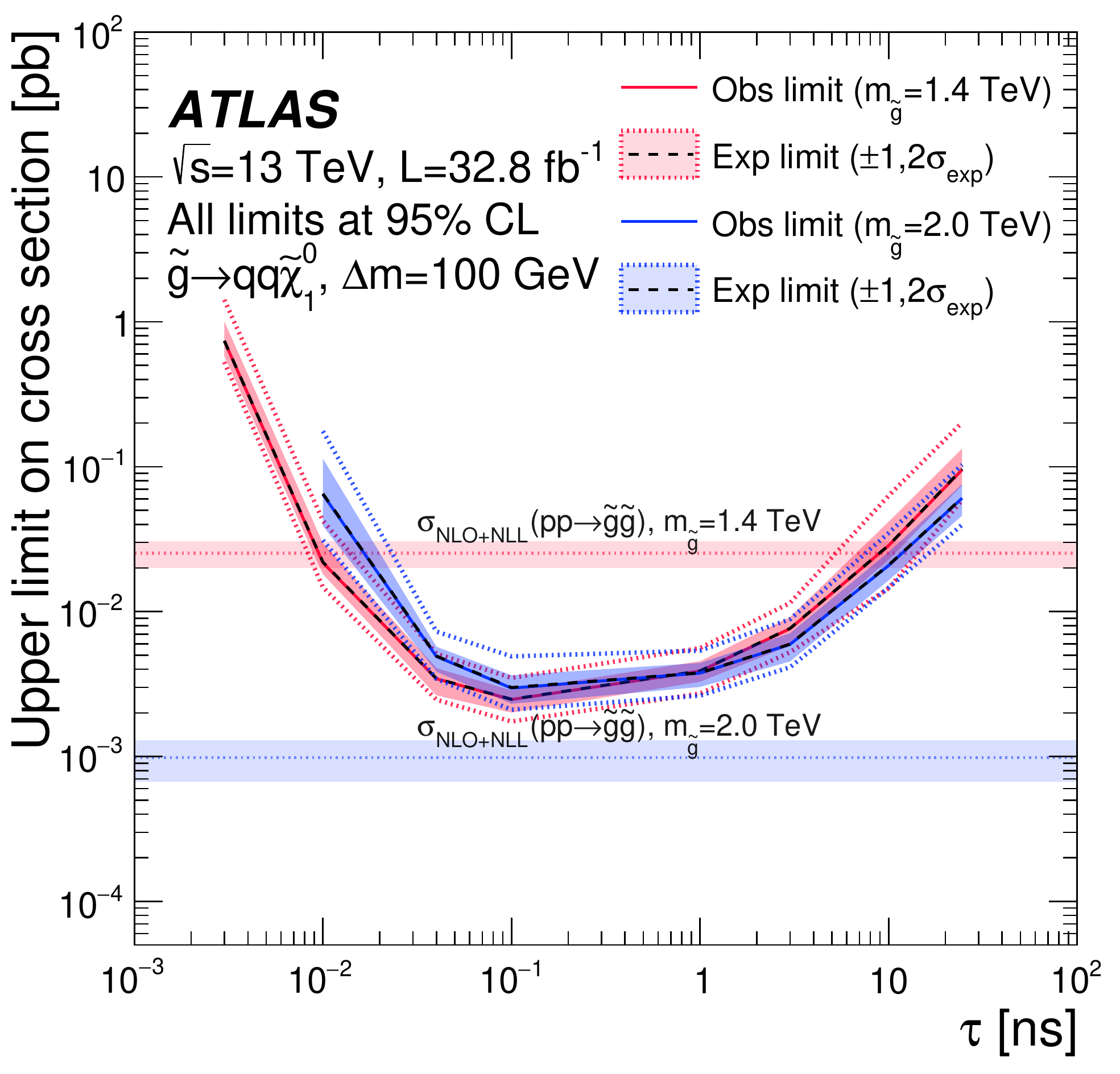}
}
\subfloat[Lower limits on $m_{\gluino}$]{
	\label{fig:mg_vs_ctau_dm_100}
	\includegraphics[width=0.47\textwidth]{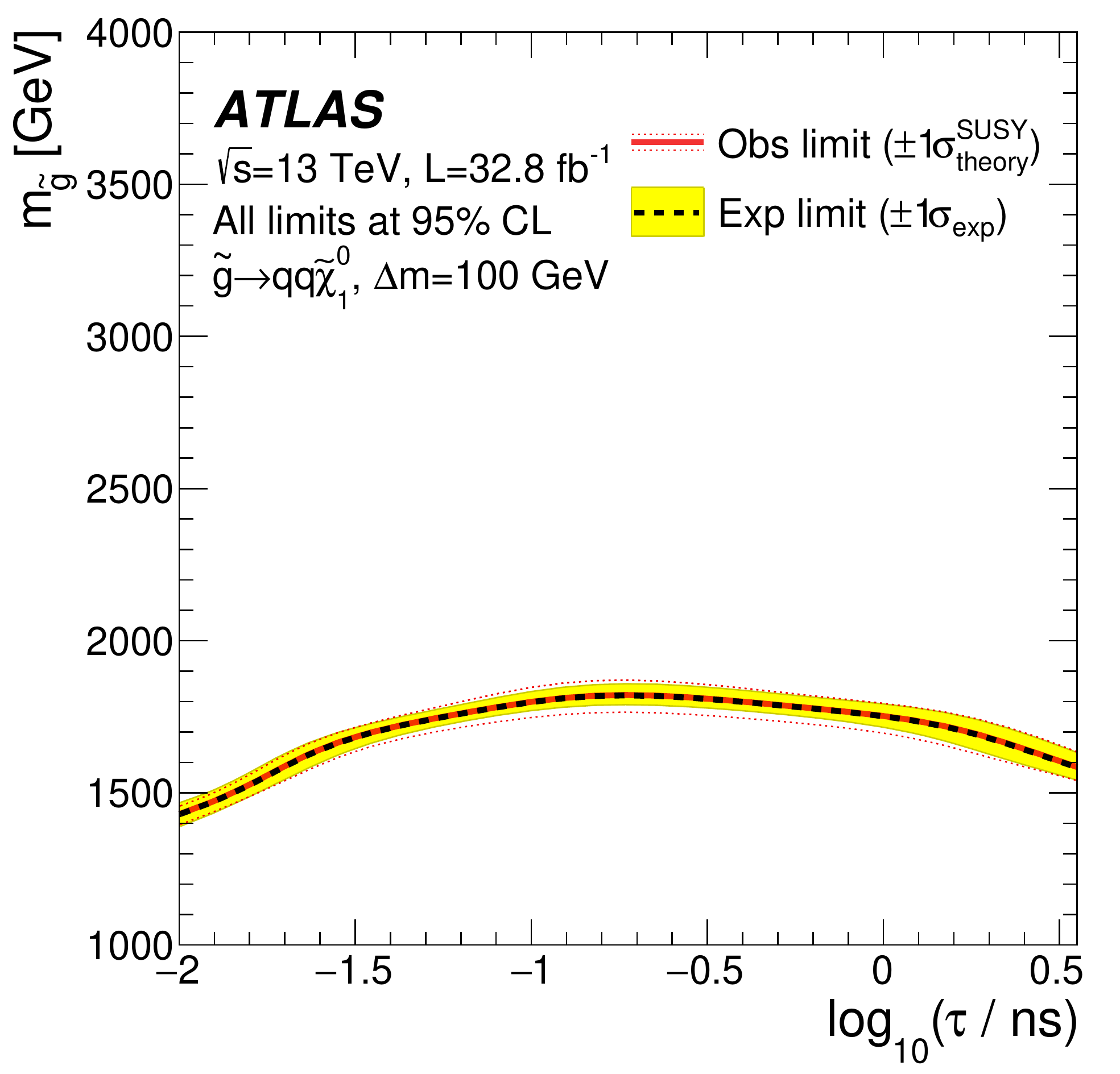}
}
\caption{\label{fig:exclusion_dm100} Upper 95\% CL limits on the signal cross section are shown in \protect\subref{fig:upperlimit_dm_100} for $m_{\gluino}=1400$~\GeV\ and $m_{\gluino}=2000$~\GeV\ as a function of lifetime $\tau$, for fixed $\Delta m=100$~\GeV. Horizontal lines denote the $\gluino\gluino$ production cross section for the same values of $m_{\gluino}$, shown with uncertainties given by variations of the renormalization and factorization scale and PDF uncertainties. The lower limit on $m_{\gluino}$ for fixed $\Delta m=100$~\GeV\ as a function of lifetime $\tau$ is shown in \protect\subref{fig:mg_vs_ctau_dm_100}. The nominal expected and observed limit contours coincide due to the signal region yield's high level of agreement with expectation.}
\end{center}
\end{figure}

\begin{figure}[!htbp]
\begin{center}
\subfloat[Upper limits on production cross section]{
	\label{fig:upperlimit_tau_1}
	\includegraphics[width=0.47\textwidth]{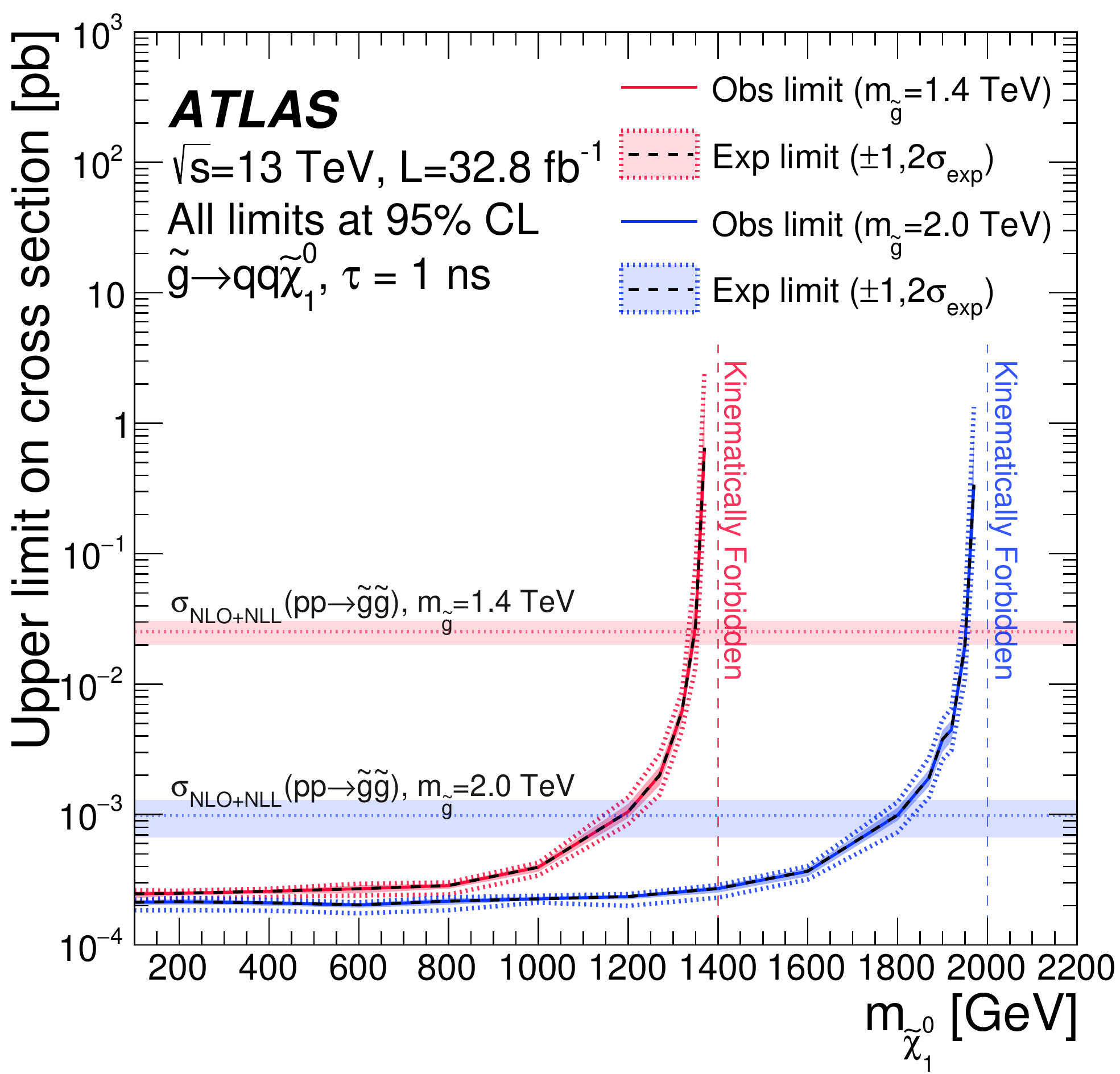}
}
\subfloat[Lower limits on $m_{\gluino}$ and $m_{\tilde{\chi}_{1}^{0}}$]{
	\label{fig:mlsp_vs_mg_tau_1}
	\includegraphics[width=0.47\textwidth]{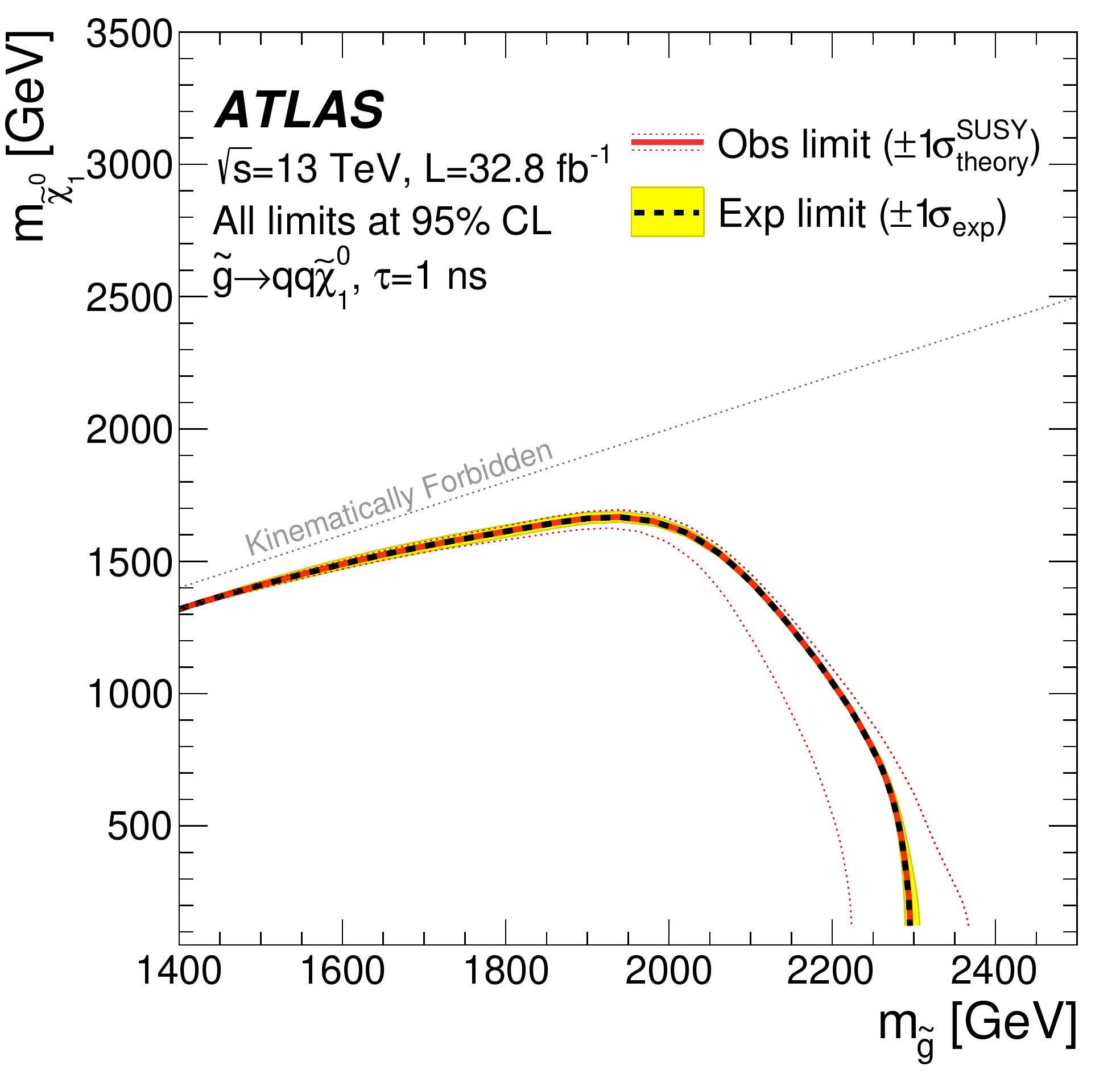}
}
\caption{\label{fig:exclusion_tau1} Upper 95\% CL limits on the signal cross section are shown in \protect\subref{fig:upperlimit_tau_1} for $m_{\gluino}=1400$~\GeV\ and $m_{\gluino}=2000$~\GeV\ as a function of $m_{\tilde{\chi}_{1}^{0}}$, for fixed $\tau=1$~ns. Horizontal lines denote the $\gluino\gluino$ production cross section for the same values of $m_{\gluino}$, shown with uncertainties given by variations of the renormalization and factorization scale and PDF uncertainties. The 95\% CL limit as a function of $m_{\gluino}$ and $m_{\tilde{\chi}_{1}^{0}}$ is shown in \protect\subref{fig:mlsp_vs_mg_tau_1} for fixed $\tau=1$~ns. The nominal expected and observed limit contours coincide due to the signal region yield's high level of agreement with expectation.}
\end{center}
\end{figure}

In the absence of a statistically significant excess in the data, exclusion limits are placed on $R$-hadron models. These 95\% confidence-level (CL) upper limits are calculated following the CL$_{\text{s}}$ prescription~\cite{Read:2002hq} with the profile likelihood used as the test statistic, using the HistFitter~\cite{Baak:2014wma} framework with pseudo-experiments.  Upper limits on the cross section for gluino pair-production as a function of gluino lifetime are shown in \Fig{\ref{fig:exclusion_mlsp100}} for example values of $m_{\gluino}$ and $m_{\tilde{\chi}_{1}^{0}}=100$~\GeV. Also shown are the signal production cross sections for these gluino masses. Reduced signal selection efficiencies for low-\Deltam\ samples result in less stringent cross-section limits. For $\Delta m = 100$~\GeV, the limits are shown in \Fig{\ref{fig:exclusion_dm100}}. Lower limits on the gluino mass are also shown as a function of gluino lifetime in \Figs{\ref{fig:exclusion_mlsp100}}{\ref{fig:exclusion_dm100}}. DV-level fiducial volume and PV-distance requirements reduce the exclusion power in the high and low extremes of gluino lifetime. Similarly, for a fixed gluino lifetime of $\tau=1$~ns, 95\% CL exclusion curves are shown as a function of $m_{\gluino}$ and $m_{\tilde{\chi}_{1}^{0}}$ in \Fig{\ref{fig:exclusion_tau1}}. For $m_{\tilde{\chi}_{1}^{0}}=100$~\GeV, gluino masses are excluded below $2.29$~\TeV\ at $\tau=1$~ns and below $2.37$~\TeV\ at around $\tau=0.17$~ns.

\FloatBarrier

\section{Conclusions}
\label{sec:conclusion}

A search for massive, long-lived particles with decays giving rise to displaced multi-track vertices is performed with \lumi\ of $pp$ collisions at $\sqrt{s} = 13$~\TeV\ collected by the ATLAS experiment at the LHC. The search presented is sensitive to models predicting events with significant $\MET$ and at least one displaced vertex with five or more tracks and a visible invariant mass greater than $10$~\GeV. With an expected background of $0.02^{+0.02}_{-0.01}$ events, no events in the data sample were observed in the signal region. With results consistent with the background-only hypothesis, exclusion limits are derived for models predicting the existence of such particles, reaching roughly $m_{\gluino} = 2000$~\GeV\ to $2370$~\GeV\ for $m_{\tilde{\chi}_{1}^{0}}=100$~\GeV\ and gluino lifetimes between $0.02$ and $10$~ns.
For a fixed gluino--neutralino mass difference of $\Delta m=100$~\GeV, exclusion limits reach roughly $m_{\gluino} = 1550$~\GeV\ to $1820$~\GeV\ for gluino lifetimes between $0.02$ and $4$~ns.

\section*{Acknowledgements}



We thank CERN for the very successful operation of the LHC, as well as the
support staff from our institutions without whom ATLAS could not be
operated efficiently.

We acknowledge the support of ANPCyT, Argentina; YerPhI, Armenia; ARC, Australia; BMWFW and FWF, Austria; ANAS, Azerbaijan; SSTC, Belarus; CNPq and FAPESP, Brazil; NSERC, NRC and CFI, Canada; CERN; CONICYT, Chile; CAS, MOST and NSFC, China; COLCIENCIAS, Colombia; MSMT CR, MPO CR and VSC CR, Czech Republic; DNRF and DNSRC, Denmark; IN2P3-CNRS, CEA-DSM/IRFU, France; SRNSF, Georgia; BMBF, HGF, and MPG, Germany; GSRT, Greece; RGC, Hong Kong SAR, China; ISF, I-CORE and Benoziyo Center, Israel; INFN, Italy; MEXT and JSPS, Japan; CNRST, Morocco; NWO, Netherlands; RCN, Norway; MNiSW and NCN, Poland; FCT, Portugal; MNE/IFA, Romania; MES of Russia and NRC KI, Russian Federation; JINR; MESTD, Serbia; MSSR, Slovakia; ARRS and MIZ\v{S}, Slovenia; DST/NRF, South Africa; MINECO, Spain; SRC and Wallenberg Foundation, Sweden; SERI, SNSF and Cantons of Bern and Geneva, Switzerland; MOST, Taiwan; TAEK, Turkey; STFC, United Kingdom; DOE and NSF, United States of America. In addition, individual groups and members have received support from BCKDF, the Canada Council, CANARIE, CRC, Compute Canada, FQRNT, and the Ontario Innovation Trust, Canada; EPLANET, ERC, ERDF, FP7, Horizon 2020 and Marie Sk{\l}odowska-Curie Actions, European Union; Investissements d'Avenir Labex and Idex, ANR, R{\'e}gion Auvergne and Fondation Partager le Savoir, France; DFG and AvH Foundation, Germany; Herakleitos, Thales and Aristeia programmes co-financed by EU-ESF and the Greek NSRF; BSF, GIF and Minerva, Israel; BRF, Norway; CERCA Programme Generalitat de Catalunya, Generalitat Valenciana, Spain; the Royal Society and Leverhulme Trust, United Kingdom.

The crucial computing support from all WLCG partners is acknowledged gratefully, in particular from CERN, the ATLAS Tier-1 facilities at TRIUMF (Canada), NDGF (Denmark, Norway, Sweden), CC-IN2P3 (France), KIT/GridKA (Germany), INFN-CNAF (Italy), NL-T1 (Netherlands), PIC (Spain), ASGC (Taiwan), RAL (UK) and BNL (USA), the Tier-2 facilities worldwide and large non-WLCG resource providers. Major contributors of computing resources are listed in Ref.~\cite{ATL-GEN-PUB-2016-002}.


\printbibliography

\newpage 
\begin{flushleft}
{\Large The ATLAS Collaboration}

\bigskip

M.~Aaboud$^\textrm{\scriptsize 137d}$,
G.~Aad$^\textrm{\scriptsize 88}$,
B.~Abbott$^\textrm{\scriptsize 115}$,
O.~Abdinov$^\textrm{\scriptsize 12}$$^{,*}$,
B.~Abeloos$^\textrm{\scriptsize 119}$,
S.H.~Abidi$^\textrm{\scriptsize 161}$,
O.S.~AbouZeid$^\textrm{\scriptsize 139}$,
N.L.~Abraham$^\textrm{\scriptsize 151}$,
H.~Abramowicz$^\textrm{\scriptsize 155}$,
H.~Abreu$^\textrm{\scriptsize 154}$
% ...
\bigskip
\\
% ...
$^{12}$ Institute of Physics, Azerbaijan Academy of Sciences, Baku, Azerbaijan\\
% ...
$^{88}$ CPPM, Aix-Marseille Universit{\'e} and CNRS/IN2P3, Marseille, France\\
% ...
$^{115}$ Homer L. Dodge Department of Physics and Astronomy, University of Oklahoma, Norman OK, United States of America\\
% ...
$^{119}$ LAL, Univ. Paris-Sud, CNRS/IN2P3, Universit{\'e} Paris-Saclay, Orsay, France\\
% ...
$^{137}$ $^{(a)}$ Facult{\'e} des Sciences Ain Chock, R{\'e}seau Universitaire de Physique des Hautes Energies - Universit{\'e} Hassan II, Casablanca; $^{(b)}$ Centre National de l'Energie des Sciences Techniques Nucleaires, Rabat; $^{(c)}$ Facult{\'e} des Sciences Semlalia, Universit{\'e} Cadi Ayyad, LPHEA-Marrakech; $^{(d)}$ Facult{\'e} des Sciences, Universit{\'e} Mohamed Premier and LPTPM, Oujda; $^{(e)}$ Facult{\'e} des sciences, Universit{\'e} Mohammed V, Rabat, Morocco\\
% ...
$^{139}$ Santa Cruz Institute for Particle Physics, University of California Santa Cruz, Santa Cruz CA, United States of America\\
% ...
$^{151}$ Department of Physics and Astronomy, University of Sussex, Brighton, United Kingdom\\
% ...
$^{154}$ Department of Physics, Technion: Israel Institute of Technology, Haifa, Israel\\
$^{155}$ Raymond and Beverly Sackler School of Physics and Astronomy, Tel Aviv University, Tel Aviv, Israel\\
% ...
$^{161}$ Department of Physics, University of Toronto, Toronto ON, Canada\\
% ...
$^{a}$ Also at Department of Physics, King's College London, London, United Kingdom\\
$^{b}$ Also at Institute of Physics, Azerbaijan Academy of Sciences, Baku, Azerbaijan\\
$^{c}$ Also at Novosibirsk State University, Novosibirsk, Russia\\
$^{d}$ Also at TRIUMF, Vancouver BC, Canada\\
$^{e}$ Also at Department of Physics {\&} Astronomy, University of Louisville, Louisville, KY, United States of America\\
% ...
$^{*}$ Deceased
\end{flushleft}




\end{document}